\documentclass[aps,pra,showpacs,floatfix]{revtex4}
\usepackage{graphicx}
\usepackage{times}
\usepackage{nicefrac}
\usepackage{amsmath}
\usepackage{amsfonts}
\usepackage{amssymb}
\usepackage{amsthm}
\usepackage{epsf}
\usepackage{bm}
\usepackage{bbm}

\usepackage{dcolumn}
\newcolumntype{.}{D{x}{}{-1}}

\newcommand{\be}{\begin{eqnarray}}
\newcommand{\ee}{\end{eqnarray}}
\newcommand{\la}{\langle}
\newcommand{\ra}{\rangle}

\newcommand{\eps}{\epsilon}
\newcommand{\veps}{\varepsilon}

\newcommand{\balpha}{\bm{\alpha}}

\newcommand{\bfk}{{\bf k}}

\newcommand{\bfx}{{\bf x}}
\newcommand{\bfy}{{\bf y}}

\begin{document}
\title{Interelectronic-interaction effects on the two-photon decay rates
of heavy He-like ions}
\author{A. V. Volotka,$^{1,2,3,4}$ A. Surzhykov,$^{1,2}$ V. M. Shabaev,$^{4}$
and G. Plunien$^{3}$}

\affiliation{
$^1$Physikalisches Institut, Universit\"at Heidelberg, D-69120 Heidelberg, Germany\\
$^2$GSI Helmholtzzentrum f\"ur Schwerionenforschung, D-64291 Darmstadt, Germany\\
$^3$Institut f\"ur Theoretische Physik, Technische Universit\"at Dresden,
Mommsenstrasse 13, D-01062 Dresden, Germany\\
$^4$Department of Physics, St. Petersburg State University,
Oulianovskaya 1, Petrodvorets, 198504 St. Petersburg, Russia\\
}

\begin{abstract}
Based on a rigorous QED approach a theoretical analysis is performed for the two-photon
transitions in heavy He-like ions. Special attention is paid to the
interelectronic-interaction corrections to the decay rates that are taken into
account within the two-time Green-function method. Detailed calculations are carried
out for the two-photon transitions $2^1S_0 \rightarrow 1^1S_0$ and
$2^3S_1 \rightarrow 1^1S_0$ in He-like ions within the range of nuclear
numbers $Z = 28 - 92$. The total decay rates together with the spectral distributions
are given. The obtained results are compared with experimental values and previous
calculations.
\end{abstract}

\maketitle

\section{Introduction}
\label{sec:1}

The two-photon process involving simultaneous emission of two photons was
theoretically predicted by G\"oppert-Mayer in 1931 \cite{goeppert:1931:273}.
It arises from a second-order interaction between an atom and the electromagnetic
field resulting in sharing the transition energy between the two photons.
The energy distribution of the two-photon spontaneous emission forms
a continuous spectrum in contrast to the one-photon process, where
the photon frequency equals to the transition energy. Various characteristics
of the two-photon transitions, such as total and energy-differential decay rates,
angular and polarization correlations of the emitted photons were widely investigated
for heavy hydrogenlike ions (see, e.g., Refs.~\cite{drake:1986:2871,santos:1998:43,
surzhykov:2005:022509,labzowsky:2005:265,amaro:2009:062504}). Due to the recent
advances in the experimental technique heavy He-like ions became promising candidates for
studying the two-photon decays in the high-$Z$ domain. Here the $2^1S_0$ state is of
special interest, since this state primarily decays into the ground state via
two-photon emission. The first theoretical two-photon decay rate of the $2^1S_0$ state
in helium was presented by Dalgarno \cite{dalgarno:1966:311}.
Later accurate nonrelativistic calculations, including the estimation of the
relativistic effects, of the two-photon transition rates $2^1S_0 \rightarrow 1^1S_0 + 2\gamma(\rm E1)$
for He-like ions were performed by Drake \cite{drake:1986:2871}.
The two-photon decay $2^3S_1 \rightarrow 1^1S_0 + 2\gamma(\rm E1)$ was
investigated theoretically as well \cite{bely:1969:37,drake:1969:25},
although its rates are smaller than the corresponding one-photon M1 rates by
a factor of about $10^{-4}$. Up to date the most accurate fully relativistic
calculations of the two-photon decay rates of the $2^1S_0$ and $2^3S_1$ states
in the highly charged ions were performed using relativistic
configuration-interaction wave functions in Ref.~\cite{derevianko:1997:1288}.
Apart from the total and energy-differential decay rates the angular correlations
in the two-photon decay of He-like ions have also been investigated recently
\cite{surzhykov:2010:042510}.

The lifetimes of metastable $2^1S_0$ level in He-like ions have been measured up
to $Z = 41$. The most precise measurements have been made in Kr$^{34+}$
\cite{marrus:1683:1986}, Br$^{33+}$ \cite{dunford:1993:1929}, and Ni$^{26+}$
\cite{dunford:1993:2729} where uncertainties of about 1\% have been reported.
However, till present the two-photon decay of the $2^3S_1$ level in He-like ions has
not been observed. As opposed to the total decay rate measurements, the observation
of the energy-differential spectrum carries more detailed information about the
entire atomic structure. Several experimental efforts have been made during the last
two decades to accurately determine the spectral shape of the two-photon distribution for
$2^1S_0$ decay in He-like ions \cite{mokler:1990:3108,ali:1997:994,schaeffer:1999:489}.
The cleanest spectrum has been obtained recently in
Refs.~\cite{kumar:2009:19,trotsenko:2010:033001}, unambiguously confirmed predictions
of relativistic many-body theory as compared to the nonrelativistic calculations.

Since the two electrons in He-like ions are strongly correlated,
it is important to take into account the interelectronic-interaction effects
when studying the two-photon decays.
In previous calculations the correlation effects were accounted for by means
of nonrelativistic Hylleraas variational wave functions \cite{drake:1986:2871},
relativistic configuration-interaction (CI) wave functions \cite{derevianko:1997:1288},
or by means of relativistic wave functions in screening potentials
\cite{savukov:2002:062507,trotsenko:2010:033001,surzhykov:2010:042510,shabaev:2010:052102}.
However, a rigorous description of high-$Z$ systems requires the quantum electrodynamic
(QED) approach, which treats systematically radiative and correlation corrections
order by order. Future progress in the experimental techniques will allow to observe
QED corrections to the transition amplitudes. In particular, recent precise measurements
of the one-photon decay rates of the $(1s^2 2s^2 2p)\,^2P_{3/2}$ state in B-like Ar
\cite{lapierre:2005:183001,lapierre:2006:052507} have been shown to be sensitive to the
one- and many-electron QED effects \cite{tupitsyn:2005:062503,volotka:2006:293,
volotka:2008:167}. The QED treatment of the correlation effects differs
from the many-body perturbation theory by the frequency-dependent contribution.
The first QED evaluation of the interelectronic-interaction correction of first order
in $1/Z$ to the one-photon decay rates was performed in
Ref.~\cite{indelicato:2004:062506} employing the two-time Green-function
method \cite{shabaev:1990:43,shabaev:1990:83,shabaev:2002:119}; later these calculations
were confirmed in Ref.~\cite{andreev:2009:032515} by means of the line profile approach
\cite{andreev:2008:135}. The main goals of the present paper are the derivation of formulas
for the interelectronic-interaction corrections to the two-photon decays from the first principles
of QED and the numerical evaluations of the two-photon transitions
$2^1S_0 \rightarrow 1^1S_0$ and $2^3S_1 \rightarrow 1^1S_0$ in the He-like ions.
The paper is organized as follows: In the next section the process of the two-photon
emission is described in the framework of the two-time Green-function method.
The calculation formulas for the first-order interelectronic-interaction corrections
to the two-photon transition amplitude are derived starting in the zeroth-order
approximation with the Coulomb potential of the nucleus and with a local screening
potential. In Sec.~\ref{sec:3} we present the numerical results
for the two-photon decay rates of $2^1S_0$ and $2^3S_1$ states in He-like ions.
Beyond the dominant channel of the emission of two electric-dipole (E1) photons
the higher multipoles contributions are also taken into account. The total and
energy-differential decay rates are presented within the range of nuclear numbers
$Z = 28 - 92$. Comparison with previous theoretical calculations and with
experiment are given. We close with a short summary, where we point out the
main achievements of the present work.

Relativistic units ($\hbar = 1,\,c = 1,\,m = 1$) and the Heaviside charge
unit [$\alpha = e^2/(4\pi)$, $e<0$] are used throughout the paper.

\section{Basic formulas}
\label{sec:2}

According to the basic principles of QED \cite{berestetsky}, the transition
probability from the electronic state $A$ to $B$ accompanied by emission of two photons 
with wave vectors $k_{f_1}$, $k_{f_2}$ and polarizations $\eps_{f_1}$, $\eps_{f_2}$,
respectively, is given by
\be
dW_{B;A}(k_{f_1},\eps_{f_1},k_{f_2},\eps_{f_2}) =
  2\pi |\tau_{\gamma_{f_1},\gamma_{f_2},B;A}|^2\delta(E_B+k_{f_1}^0+k_{f_2}^0-E_A)
  d\bfk_{f_1}d\bfk_{f_2}\,,
\ee
where $\tau_{\gamma_{f_1},\gamma_{f_2},B;A}$ is the transition amplitude
which is related to the $S$-matrix element by
\be \label{deftau}
S_{\gamma_{f_1},\gamma_{f_2},B;A} =
  \la k_{f_1},\eps_{f_1},k_{f_2},\eps_{f_2};B|\hat{S}|A\ra = 
  2\pi i\,\tau_{\gamma_{f_1},\gamma_{f_2},B;A}\,\delta(E_B+k_{f_1}^0+k_{f_2}^0-E_A)\,,
\ee
$E_A$ and $E_B$ are the energies of the initial state $A$
and the final state $B$, respectively.
According to the standard reduction technique, the $S$-matrix element
can be written as
\be \label{s1}
S_{\gamma_{f_1},\gamma_{f_2},B;A} = -Z_{3}^{-1} \int d^4 y_1 d^4 y_2
  \frac{\eps_{f_1}^{\nu_1 *} \, e^{ik_{f_1}\cdot y_1}}{\sqrt{2k_{f_1}^{0}(2\pi )^3}}
  \frac{\eps_{f_2}^{\nu_2 *} \, e^{ik_{f_2}\cdot y_2}}{\sqrt{2k_{f_2}^{0}(2\pi )^3}}
  \la B|T j_{\nu_1}(y_1) j_{\nu_2}(y_2)|A\ra\,,
\ee
where $j_{\nu}(y)=(e/2)[\overline{\psi}(y)\gamma_\nu,\psi(y)]$ is the Dirac
current density operator and $Z_3$ is a renormalization constant for
the emitted photons lines \cite{itzykson}.
Here the electron-positron current operator $j_{\nu}(y)$ as well as
the initial and final state vectors are given in the Heisenberg picture.
Eq.~(\ref{s1}) can be written as
\be \label{s2}
S_{\gamma_{f_1},\gamma_{f_2},B;A} &=& -2\pi Z_{3}^{-1} \delta(E_B+k_{f_1}^0+k_{f_2}^0-E_A)
  \int d\bfy_1 d\bfy_2 A_{f_1}^{\nu_1 *}(\bfy_1) A_{f_2}^{\nu_2 *}(\bfy_2)
  \int_{-\infty}^{\infty} dt\,e^{i k_{f_1}^0 t}\,  
  \la B|T j_{\nu_1}(t,\bfy_1) j_{\nu_2}(0,\bfy_2)|A\ra\nonumber\\
  &=&-2\pi Z_{3}^{-1} \delta(E_B+k_{f_1}^0+k_{f_2}^0-E_A)
  \int d\bfy_1 d\bfy_2 A_{f_1}^{\nu_1 *}(\bfy_1) A_{f_2}^{\nu_2 *}(\bfy_2)
  \left[\int_0^{\infty} dt\,e^{i k_{f_1}^0 t}\,  
  \la B|j_{\nu_1}(t,\bfy_1) j_{\nu_2}(0,\bfy_2)|A\ra\right.\nonumber\\
  &+&\left.\int_{-\infty}^0 dt\,e^{i k_{f_1}^0 t}\,  
  \la B|j_{\nu_2}(0,\bfy_2) j_{\nu_1}(t,\bfy_1)|A\ra\right]
\,,
\ee
where
\be \label{defA}
A_{f}^{\nu}(\bfx) = \frac{{\eps}_{f}^{\nu} \, e^{i{\bf k}_{f} \cdot \bfx}}
  {\sqrt{2k_{f}^{0}(2\pi )^3}}
\ee
is the wave function of the emitted photon.

In order to evaluate this $S$-matrix element the information about the entire atomic
structure is needed. This information is contained in the Green functions.
To obtain this information and to formulate perturbation theory we employ
the two-time Green-function method \cite{shabaev:1990:43,shabaev:1990:83,shabaev:2002:119}.
We introduce the following Green function to describe the process of a two-photon
emission by an $N$-electron ion
\be
\lefteqn{ {\cal G}_{\gamma_{f_1},\gamma_{f_2}}(E',E,k_{f_1}^0;
  {\bfx}_1^{\prime},\dots,{\bf x}_N^{\prime};{\bf x}_1,\dots,{\bf x}_N)
  \delta(E'+k_{f_1}^0+k_{f_2}^0-E)}\nonumber\\
  &=&\left(\frac{i}{2\pi}\right)^2\frac{1}{N!}
  \int_{-\infty}^{\infty} dx^0 dx'^0 \int d^4 y_1 d^4 y_2
  \,e^{iE^{\prime}x'^0-iEx^0+ik_{f_1}^0y_1^0+ik_{f_2}^0y_2^0}\,
  A_{f_1}^{\nu_1 *}(\bfy_1) A_{f_2}^{\nu_2 *}(\bfy_2)\nonumber\\
  &\times& \la 0|T\psi (x'^0,{\bf x}_1^{\prime})\dots\psi (x'^0,{\bf x}_N^{\prime})
  j_{\nu_1}(y_1)j_{\nu_2}(y_2)\overline{\psi} (x^0,{\bf x}_N)\cdots\overline{\psi} (x^0,{\bf x}_1)|0\ra\,,
\ee
where $\psi(x)$ is the electron-positron field operator in the Heisenberg representation.
In a general case, we imply that to zeroth approximation the vector $A$ belongs to
the $s_A$-dimensional subspace $\Omega_A$ of degenerate (or quasi-degenerate) states,
and the state $B$ belongs to the $s_B$-dimensional subspace $\Omega_B$. $P^{(0)}_A$ and
$P^{(0)}_B$ are the projectors onto the corresponding subspaces, 
\be
P_A^{(0)} = \sum_{k_A=1}^{s_A} u_{k_A} u^\dagger_{k_A}\,,\hspace{1cm}
P_B^{(0)} = \sum_{k_B=1}^{s_B} u_{k_B} u^\dagger_{k_B}\,,
\ee
and $u_{k_A}$ and $u_{k_B}$ are the unperturbed states of the $N$-electron system,
constructed as linear combinations of one-determinant wave functions.
From the spectral representation we find that the Green function
${\cal G}_{\gamma_{f_1},\gamma_{f_2}}(E',E,k_{f_1}^0)$ has isolated poles in the complex planes
$E'$ and $E$, at $E'\sim E_B^{(0)}$ and $E\sim E_A^{(0)}$, in the exact
energies $E'=E_{k_B}$ and $E'=E_{k_A}$,
respectively,
\be \label{calgreen}
\lefteqn{ {\cal G}_{\gamma_{f_1},\gamma_{f_2}}(E',E,k_{f_1}^0;
  {\bfx}_1^{\prime},\dots,{\bf x}_N^{\prime};{\bf x}_1,\dots,{\bf x}_N)
  \delta(E'+k_{f_1}^0+k_{f_2}^0-E)}\nonumber\\
  &=&\frac{1}{2\pi}\frac{1}{N!} \sum_{k_A=1}^{s_A} \sum_{k_B=1}^{s_B}
  \frac{1}{E'-E_{k_B}}\frac{1}{E-E_{k_A}}
  \int d\bfy_1 d\bfy_2 A_{f_1}^{\nu_1 *}(\bfy_1) A_{f_2}^{\nu_2 *}(\bfy_2)
  \la 0|\psi (0,{\bf x}_1^{\prime})\dots\psi (0,{\bf x}_N^{\prime})|k_B \ra\nonumber\\
  &\times&\left[\int_0^\infty dt\,e^{iE't-iE_{k_B}t+ik_{f_1}^0t}\,
  \la k_B|j_{\nu_1}(t,\bfy_1)j_{\nu_2}(0,\bfy_2)|k_A \ra
  +\int_{-\infty}^0 dt\,e^{iE_{k_A}t-iEt+ik_{f_1}^0t}\,
  \la k_B|j_{\nu_2}(0,\bfy_2)j_{\nu_1}(t,\bfy_1)|k_A \ra\right]\nonumber\\
  &\times& \la k_A|\overline{\psi} (0,{\bf x}_N)\dots\overline{\psi}(0,{\bf x}_1)|0\ra
  +\mbox{terms regular at $E'\sim E_B^{(0)}$ or $E\sim E_A^{(0)}$}\,,
\ee
where $|k_A\ra$ and $|k_B\ra$ denote the states corresponded to the exact energies
$E_{k_A}$ and $E_{k_B}$ from the subspaces $\Omega_A$ and $\Omega_B$, respectively.
Let us now project this Green function on the subspace of initial ($\Omega_A$)
and final ($\Omega_B$) states
\be \label{smallgreen}
g_{\gamma_{f_1},\gamma_{f_2},B;A}(E',E,k_{f_1}^0) = P_B^{(0)}
  {\cal G}_{\gamma_{f_1},\gamma_{f_2}}(E',E,k_{f_1}^0) \gamma_1^0 \dots \gamma_N^0 P_A^{(0)}\,.
\ee
Comparing Eq.~(\ref{s2}) with Eq.~(\ref{calgreen}) and taking
into account the definition (\ref{smallgreen}), we obtain
\be \label{s3}
S_{\gamma_{f_1},\gamma_{f_2},B;A} = Z_3^{-1}\delta(E_B+k_{f_1}^0+k_{f_2}^0-E_A)
  \oint_{\Gamma_B}dE'\oint_{\Gamma_A}dE\,v_B^{\dag}\,
  g_{\gamma_{f_1},\gamma_{f_2},B;A}(E',E,k_{f_1}^0)\,v_A\,,
\ee
where $v_A$ and $v_B$ are solutions of a generalized eigenvalue problem in the
degenerate subspaces of the initial and final states, respectively
(see Ref.~\cite{shabaev:2002:119} for details), the contours $\Gamma_{A}$ and
$\Gamma_{B}$ enclose the poles corresponding to the initial and final levels,
respectively, and exclude all other singularities of Green function
$g_{\gamma_{f_1},\gamma_{f_2},B;A}$. Eq.~(\ref{s3}) represents the general
relation between the $S$-matrix element of the two-photon transition and
the two-time Green functions.

Further we consider the single initial and final states. In this case, the vectors
$v_A$ and $v_B$ simply appear as normalization factors and the $S$-matrix element
can be written as
\be \label{s4}
S_{\gamma_{f_1},\gamma_{f_2},B;A} &=& Z_3^{-1}\delta(E_B+k_{f_1}^0+k_{f_2}^0-E_A)
  \oint_{\Gamma_B}dE'\oint_{\Gamma_A}dE\,g_{\gamma_{f_1},\gamma_{f_2},B;A}(E',E,k_{f_1}^0)\nonumber\\
  &\times&\Bigl[\frac{1}{2\pi i}\oint_{\Gamma_B}dE\,g_{BB}(E)\Bigr]^{-1/2}
  \Bigl[\frac{1}{2\pi i}\oint_{\Gamma_A}dE\,g_{AA}(E)\Bigr]^{-1/2}\,,
\ee
where the Green functions $g_{AA}$ and $g_{BB}$ are defined by
\be
g_{AA}(E) = \la u_A | {\cal G}(E) \gamma^{0}_1\dots\gamma^{0}_N | u_A \ra\,,\hspace{1cm}
g_{BB}(E) = \la u_B | {\cal G}(E) \gamma^{0}_1\dots\gamma^{0}_N | u_B \ra\,,
\ee
with
\be
\lefteqn{ {\cal G}(E;{\bf x}_1^{\prime},\dots,{\bf x}_N^{\prime};{\bf x}_1,\dots,{\bf x}_N)
  \delta (E-E^{\prime}) }\nonumber\\
  &=&\frac{1}{2\pi i}\frac{1}{N!}\int_{-\infty}^{\infty}dx^0 dx'^0\,e^{iE^{\prime}x'^0-iEx^0}
  \la 0|T\psi (x'^0,{\bf x}_1^{\prime})\dots\psi(x'^0,{\bf x}_N^{\prime})
  \overline{\psi}(x^0,{\bf x}_N)\dots\overline{\psi}(x^0,{\bf x}_1)|0\ra\,.
\ee
The Green function ${\cal G}(E)$ contains the complete information about the
energy levels of the ion \cite{shabaev:2002:119}.
The $S$-matrix element $S_{\gamma_{f_1},\gamma_{f_2},B;A}$ expressed in terms
of the two-time Green functions $g_{\gamma_{f_1},\gamma_{f_2},B;A}$, $g_{AA}$, and $g_{BB}$
via Eq.~(\ref{s4}) can be calculated order by order by applying perturbation theory
to the Green functions. The Feynman rules for the Green functions are given in
Ref.~\cite{shabaev:2002:119}.

In the following we consider the two-photon transitions in He-like ions.
The zeroth-order two-electron wave functions are constructed in the $jj$-coupling scheme
as linear combinations of the Slater determinants, $A = (a_1,a_2)_{J_A M_A}$,
$B = (b_1,b_2)_{J_B M_B}$, as
\be
u_A = F_A \frac{1}{\sqrt{2}} \sum_P (-1)^P |Pa_1 Pa_2 \ra\,,
\ee
where $F_A$ denotes the shorthand notation for the summation over the Clebsch-Gordan coefficients
\be
F_A |a_1 a_2 \ra = \sum_{m_{a_1},m_{a_2}} C^{J_A M_A}_{j_{a_1} m_{a_1} j_{a_2} m_{a_2}} |a_1 a_2 \ra
  \times\Biggl\{\begin{array}{cc}
  1\,,          & a_1\neq a_2 \\
  1/\sqrt{2}\,, & a_1  =  a_2 \\
                \end{array}\,,
\ee
$J_A$ and $j_a$ are the total angular momenta of the two- and one-electron wave
functions, respectively, $M_A$ and $m_a$ its corresponding projections,
$P$ is the permutation operator, giving rise to the sign $(-1)^P$ of the permutation.
The same notations hold for the final state $B$. The one-electron wave functions are
found by solving the Dirac equation either with the Coulomb potential of the nucleus
or with a local effective potential, which partly takes into account the
interelectronic-interaction effects.

Further we consider the pure (nonresonant) two-photon decays. While the question about
cascades we leave beyond the scope of the present paper. This question was discussed
in details in Ref.~\cite{labzowsky:2009:062514} and references therein.
In the following we also assume, that the states $A$ and $B$ have at least one common
one-electron state.

\subsection{Zeroth-order approximation}
\label{sec:2A}

\begin{figure}
\includegraphics{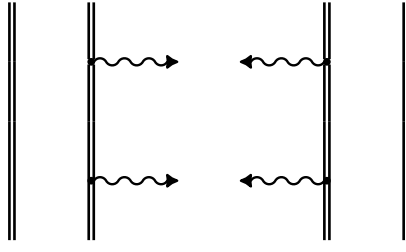}
\caption{The two-photon emission diagrams in zeroth-order approximation.
The double line indicates the electron propagators in the Coulomb field of the nucleus,
while the photon emission is depicted by the wavy line with arrow.
\label{fig:1}}
\end{figure}
In order to calculate the $S$-matrix element of the two-photon transition according
to Eq.~(\ref{s4}) we expand the two-time Green functions in perturbation series and
combine the terms of the same order. The zeroth-order two-photon transition
amplitude represented by diagrams in Fig.~\ref{fig:1} is given by
\be \label{zer1}
S^{(0)}_{\gamma_{f_1},\gamma_{f_2},B;A}=\delta(E_B+k_{f_1}^0+k_{f_2}^0-E_A)
  \oint_{\Gamma_B}dE' \oint_{\Gamma_A}dE\,
  g^{(0)}_{\gamma_{f_1},\gamma_{f_2},B;A}(E',E,k_{f_1}^0)\,,
\ee
where the superscript ``$(0)$'' indicates the order of the perturbation theory.
According to the Feynman rules we obtain
\be
\lefteqn{ g^{(0)}_{\gamma_{f_1},\gamma_{f_2},B;A}(E',E,k_{f_1}^0)
  \delta(E'+k_{f_1}^0+k_{f_2}^0-E) }\nonumber\\
  &=&
  F_A F_B \sum_{P}(-1)^P \int_{-\infty}^{\infty} dp_1^0 dp_2^0 dp_1'^0 dp_2'^0 dq^0
  \,\delta(E-p_1^0-p_2^0)\,\delta(E'-p_1'^0-p_2'^0)\nonumber\\
  &\times&
  \left\{\la Pb_2|\frac{i}{2\pi}\sum_{n_1}\frac{|n_1\ra\la n_1|}{p_2'^0-u\veps_{n_1}}
  \frac{2\pi}{i}R_{f_1}\delta(p_2'^0+k_{f_1}^0-q^0)
  \frac{i}{2\pi}\sum_{n_2}\frac{|n_2\ra\la n_2|}{q^0-u\veps_{n_2}}
  \frac{2\pi}{i}R_{f_2}\delta(q^0+k_{f_2}^0-p_2^0)\right.\nonumber\\
  &\times&
  \frac{i}{2\pi}\sum_{n_3}\frac{|n_3\ra\la n_3|}{p_2^0-u\veps_{n_3}}|a_2\ra
  \la Pb_1|\frac{i}{2\pi}\sum_{n_4}\frac{|n_4\ra\la n_4|}{p_1^0-u\veps_{n_4}}|a_1\ra
  \delta(p_1'^0-p_1^0)\nonumber\\
  &+& 
  \la Pb_1|\frac{i}{2\pi}\sum_{n_1}\frac{|n_1\ra\la n_1|}{p_1'^0-u\veps_{n_1}}
  \frac{2\pi}{i}R_{f_1}\delta(p_1'^0+k_{f_1}^0-q^0)
  \frac{i}{2\pi}\sum_{n_2}\frac{|n_2\ra\la n_2|}{q^0-u\veps_{n_2}}
  \frac{2\pi}{i}R_{f_2}\delta(q^0+k_{f_2}^0-p_1^0)\nonumber\\
  &\times&
  \left.\frac{i}{2\pi}\sum_{n_3}\frac{|n_3\ra\la n_3|}{p_1^0-u\veps_{n_3}}|a_1\ra
  \la Pb_2|\frac{i}{2\pi}\sum_{n_4}\frac{|n_4\ra\la n_4|}{p_2^0-u\veps_{n_4}}|a_2\ra
  \delta(p_2'^0-p_2^0)+(f_1 \leftrightarrow f_2) \right\}\nonumber\\
  &=&
  \frac{i}{2\pi}\frac{\delta(E'+k_{f_1}^0+k_{f_2}^0-E)}{(E'-E_B^{(0)})(E-E_A^{(0)})}
  F_A F_B \sum_{P}(-1)^P \sum_n
  \left\{\frac{\la Pb_2|R_{f_1}|n\ra\la n|R_{f_2}|a_2\ra
         \delta_{Pb_1 a_1}}{E'-\veps_{a_1}+k_{f_1}^0-\veps_n}\right.\nonumber\\
  &+&
  \left.\frac{\la Pb_1|R_{f_1}|n\ra\la n|R_{f_2}|a_1\ra
        \delta_{Pb_2 a_2}}{E'-\veps_{a_2}+k_{f_1}^0-\veps_n}
  +(f_1 \leftrightarrow f_2)\right\}\,,
\ee
where $R_f$ is the transition operator, $R_f = e\alpha_\nu A_f^{\nu *}$,
$\alpha^{\mu}=\gamma^0 \gamma^{\mu}=(1,\balpha)$, $E_A^{(0)}=\veps_{a_1}+\veps_{a_2}$
and $E_B^{(0)}=\veps_{b_1}+\veps_{b_2}$, $u = 1-i0$ preserves the proper treatment of
poles of the electron propagators, and the shorthand notation $(f_1 \leftrightarrow f_2)$
stands for the contributions with interchanged photons $f_1$ and $f_2$.
Substituting this expression into Eq.~(\ref{zer1}) and integrating over $E$ and $E'$
one obtains
\be
\label{tau_0}
\tau_{\gamma_{f_1},\gamma_{f_2},B;A}^{(0)} &=& -F_A F_B \sum_{P}(-1)^P \sum_n
  \left\{\frac{\la Pb_2|R_{f_1}|n\ra\la n|R_{f_2}|a_2\ra
         \delta_{Pb_1 a_1}}{\veps_{Pb_2}+k_{f_1}^0-\veps_n}\right.\nonumber\\
  &+&\left.\frac{\la Pb_1|R_{f_1}|n\ra\la n|R_{f_2}|a_1\ra
           \delta_{Pb_2 a_2}}{\veps_{Pb_1}+k_{f_1}^0-\veps_n}+(f_1 \leftrightarrow f_2)\right\}\,.
\ee
The corresponding differential transition probability is given by
\be
\label{dw_0}
dW_{B;A}^{(0)}(k_{f_1},\eps_{f_1},k_{f_2},\eps_{f_2}) =
  2\pi |\tau_{\gamma_{f_1},\gamma_{f_2},B;A}^{(0)}|^2
  \delta(E_B^{(0)}+k_{f_1}^0+k_{f_2}^0-E_A^{(0)}) d\bfk_{f_1}d\bfk_{f_2}\,.
\ee
Summing over the photon polarizations and integrating over the photon energies
and angles one obtains the total decay rate
\be
\label{w_0}
W_{B;A}^{(0)} = \frac{1}{2} \int_0^{\Delta^{(0)}_{AB}} dk_{f_1}^0\,
  (k_{f_1}^0)^2 (\Delta^{(0)}_{AB}-k_{f_1}^0)^2\,
  2\pi \sum_{\eps_{f_1},\eps_{f_2}} \int d\Omega_{k_{f_1}} d\Omega_{k_{f_2}}
  |\tau_{\gamma_{f_1},\gamma_{f_2},B;A}^{(0)}|^2\,,
\ee
where $\Delta^{(0)}_{AB} = E_A^{(0)} - E_B^{(0)}$.
Eqs.~(\ref{dw_0}) and (\ref{w_0}) together with Eq.~(\ref{tau_0}) describe
the zeroth-order differential and total two-photon transition probabilities,
respectively. They coincide with the corresponding formulas employed for the
calculation of the two-photon decay rates in He-like ions \cite{drake:1986:2871,
derevianko:1997:1288,labzowsky:2004:012503,surzhykov:2010:042510} in the independent
particle model approximation.
%
\subsection{First-order interelectronic-interaction correction}
\label{sec:2B}

\begin{figure}
\includegraphics{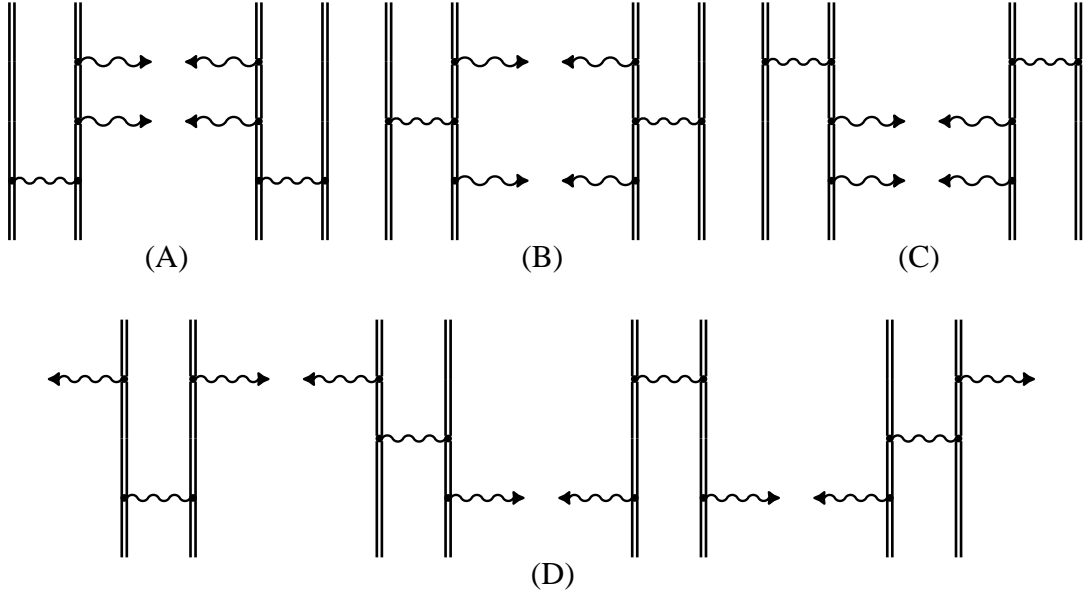}
\caption{Feynman diagrams representing the first-order interelectronic-interaction
corrections to the two-photon emission. Notations are the same as in Fig.~\ref{fig:1}.
\label{fig:2}}
\end{figure}
\begin{figure}
\includegraphics{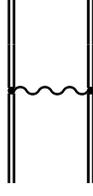}
\caption{One-photon exchange diagram. The photon propagator is represented
by the wavy line.
\label{fig:3}}
\end{figure}

With the formalism outlined above, we are ready now to derive the first-order
interelectronic-interaction corrections to the two-photon transition amplitude,
which are defined by diagrams depicted in Fig.~\ref{fig:2}.
According to Eq.~(\ref{s4}) we start from
\be \label{int1}
S^{(1)}_{\gamma_{f_1},\gamma_{f_2},B;A}&=&\delta(E_B+k_{f_1}^0+k_{f_2}^0-E_A)
  \left[\oint_{\Gamma_B}dE'\oint_{\Gamma_A}dE\,
  g^{(1)}_{\gamma_{f_1},\gamma_{f_2},B;A}(E',E,k_{f_1}^0)\right.\nonumber\\
  &-&\left.\frac{1}{2}\oint_{\Gamma_B}dE'\oint_{\Gamma_A}dE\,
  g^{(0)}_{\gamma_{f_1},\gamma_{f_2},B;A}(E',E,k_{f_1}^0)\;
  \left(\frac{1}{2\pi i}\oint_{\Gamma_A}dE\, g_{AA}^{(1)}(E)
  +\frac{1}{2\pi i}\oint_{\Gamma_B}dE\, g_{BB}^{(1)}(E)\right)\right]\,,
\ee
where $g_{AA}^{(1)}$ and $g_{BB}^{(1)}$ are defined by the first-order
interelectronic-interaction diagram depicted in Fig.~\ref{fig:3}.
Let us first consider the contribution of the diagrams shown
in Fig.~\ref{fig:2}(A). According to the Feynman rules we obtain
\be \label{int2}
\lefteqn{ g^{(1A)}_{\gamma_{f_1},\gamma_{f_2},B;A}(E',E,k_{f_1}^0)
  \delta(E'+k_{f_1}^0+k_{f_2}^0-E) }\nonumber\\
  &=&
  \left(\frac{i}{2\pi}\right)^3 F_A F_B \sum_{P}(-1)^P
  \int_{-\infty}^{\infty} dp_1^0 dp_2^0 dp_1'^0 dp_2'^0 dq_1^0 dq_2^0 d\omega
  \delta(E-p_1^0-p_2^0)\delta(E'-p_1'^0-p_2'^0)\nonumber\\
  &\times&\sum_{n_1,n_2}\left\{
  \frac{\delta(p_2'^0+k_{f_1}^0-q_1^0)\delta(q_1^0+k_{f_2}^0-q_2^0)
        \delta(q_2^0-\omega-p_2^0)\delta(p_1'^0+\omega-p_1^0)}
       {(p_1'^0-u\veps_{Pb_1})(p_2'^0-u\veps_{Pb_2})(p_1^0-u\veps_{a_1})(p_2^0-u\veps_{a_2})}\right.\nonumber\\
  &\times&
  \frac{\la Pb_2|R_{f_1}|n_1 \ra \la n_1|R_{f_2}|n_2 \ra
        \la Pb_1 n_2|I(\omega)|a_1 a_2 \ra}{(q_1^0-u\veps_{n_1})(q_2^0-u\veps_{n_2})}\nonumber\\
  &+&
  \frac{\delta(p_1'^0+k_{f_1}^0-q_1^0)\delta(q_1^0+k_{f_2}^0-q_2^0)
        \delta(q_2^0-\omega-p_1^0)\delta(p_2'^0+\omega-p_2^0)}
       {(p_1'^0-u\veps_{Pb_1})(p_2'^0-u\veps_{Pb_2})(p_1^0-u\veps_{a_1})(p_2^0-u\veps_{a_2})}\nonumber\\
  &\times&\left.
  \frac{\la Pb_1|R_{f_1}|n_1 \ra \la n_1|R_{f_2}|n_2 \ra
        \la n_2 Pb_2|I(\omega)|a_1 a_2 \ra}{(q_1^0-u\veps_{n_1})(q_2^0-u\veps_{n_2})}
  +(f_1 \leftrightarrow f_2)\right\}\,,
\ee
where $I(\omega) = e^2\alpha^{\mu}\alpha^{\nu}D_{\mu \nu}(\omega)$, and
$D_{\mu \nu}(\omega)$ is the photon propagator.
Eq.~(\ref{int2}) is conveniently divided into irreducible and reducible parts. The
reducible part is the one with $\veps_{Pb_1}+\veps_{n_2} = E_A^{(0)}$ in first term and with
$\veps_{Pb_2}+\veps_{n_2} = E_A^{(0)}$ in the second term. The irreducible part is the reminder.
Thus, we obtain for the irreducible contribution
\be
\label{int3}
\lefteqn{ g^{(1A,{\rm irr})}_{\gamma_{f_1},\gamma_{f_2},B;A}(E',E,k_{f_1}^0) }\nonumber\\
  &=&
  \left(\frac{i}{2\pi}\right)^3 F_A F_B \sum_{P}(-1)^P
  \int_{-\infty}^{\infty} dp^0 dp'^0
  \frac{1}{(E'-E_B^{(0)})(E-E_A^{(0)})}\nonumber\\
  &\times&
  \left\{\sum_{n_1,n_2}^{\veps_{Pb_1}+\veps_{n_2} \ne E_A^{(0)}}
  \left(\frac{1}{p^0-u\veps_{a_1}}+\frac{1}{E-p^0-u\veps_{a_2}}\right)
  \left(\frac{1}{p'^0-u\veps_{Pb_1}}+\frac{1}{E'-p'^0-u\veps_{Pb_2}}\right)\right.\nonumber\\
  &\times&
  \frac{\la Pb_2|R_{f_1}|n_1 \ra \la n_1|R_{f_2}|n_2 \ra
        \la Pb_1 n_2|I(p^0-p'^0)|a_1 a_2 \ra}
       {(E'-p'^0+k_{f_1}^0-u\veps_{n_1})(E-p'^0-u\veps_{n_2})}\nonumber\\
  &+&
  \sum_{n_1,n_2}^{\veps_{Pb_2}+\veps_{n_2} \ne E_A^{(0)}}
  \left(\frac{1}{p^0-u\veps_{a_2}}+\frac{1}{E-p^0-u\veps_{a_1}}\right)
  \left(\frac{1}{p'^0-u\veps_{Pb_2}}+\frac{1}{E'-p'^0-u\veps_{Pb_1}}\right)\nonumber\\
  &\times&
  \left.\frac{\la Pb_1|R_{f_1}|n_1 \ra \la n_1|R_{f_2}|n_2 \ra
        \la n_2 Pb_2|I(p^0-p'^0)|a_1 a_2 \ra}
       {(E'-p'^0+k_{f_1}^0-u\veps_{n_1})(E-p'^0-u\veps_{n_2})}
  +(f_1 \leftrightarrow f_2)\right\}\,,
\ee
and for the corresponding reducible one
\be
\label{int4}
\lefteqn{ g^{(1A,{\rm red})}_{\gamma_{f_1},\gamma_{f_2},B;A}(E',E,k_{f_1}^0) }\nonumber\\
  &=&
  \left(\frac{i}{2\pi}\right)^3 F_A F_B \sum_{P}(-1)^P
  \int_{-\infty}^{\infty} dp^0 dp'^0
  \frac{1}{(E'-E_B^{(0)})(E-E_A^{(0)})}\nonumber\\
  &\times&
  \Biggl\{\sum_{n_1,n_2}^{\veps_{Pb_1}+\veps_{n_2} = E_A^{(0)}}
  \left(\frac{1}{p^0-u\veps_{a_1}}+\frac{1}{E-p^0-u\veps_{a_2}}\right)
  \left[\frac{1}{E-E_A^{(0)}}
        \left(\frac{1}{p'^0-u\veps_{Pb_1}}+\frac{1}{E-p'^0-u\veps_{n_2}}\right)\right.\nonumber\\
  &+&
  \left.\frac{1}{(E'-p'^0-u\veps_{Pb_2})(E-p'^0-u\veps_{n_2})}\right]
  \frac{\la Pb_2|R_{f_1}|n_1 \ra \la n_1|R_{f_2}|n_2 \ra
        \la Pb_1 n_2|I(p^0-p'^0)|a_1 a_2 \ra}
       {E'-p'^0+k_{f_1}^0-u\veps_{n_1}}\nonumber\\
  &+&
  \sum_{n_1,n_2}^{\veps_{Pb_2}+\veps_{n_2} = E_A^{(0)}}
  \left(\frac{1}{p^0-u\veps_{a_2}}+\frac{1}{E-p^0-u\veps_{a_1}}\right)
  \left[\frac{1}{E-E_A^{(0)}}
        \left(\frac{1}{p'^0-u\veps_{Pb_2}}+\frac{1}{E-p'^0-u\veps_{n_2}}\right)\right.\nonumber\\
  &+&
  \left.\frac{1}{(E'-p'^0-u\veps_{Pb_1})(E-p'^0-u\veps_{n_2})}\right]
  \frac{\la Pb_1|R_{f_1}|n_1 \ra \la n_1|R_{f_2}|n_2 \ra
        \la n_2 Pb_2|I(p^0-p'^0)|a_1 a_2 \ra}
       {E'-p'^0+k_{f_1}^0-u\veps_{n_1}}\nonumber\\
  &+& (f_1 \leftrightarrow f_2)\Biggr\}\,.
\ee
The expression in curly braces of Eq.~(\ref{int3}) is a regular function of $E$ or $E'$ when
$E \approx E_A^{(0)}$ and $E' \approx E_B^{(0)}$. Substituting Eq.~(\ref{int3}) into
Eq.~(\ref{int1}) and integrating over $E$ and $E'$ we find
\be
\label{tau_A}
\tau_{\gamma_{f_1},\gamma_{f_2},B;A}^{(1A,{\rm irr})} &=& -F_A F_B \sum_{P}(-1)^P
  \left\{\sum_{n_1,n_2}^{\veps_{Pb_1}+\veps_{n_2} \ne E_A^{(0)}}
  \frac{\la Pb_2|R_{f_1}|n_1 \ra \la n_1|R_{f_2}|n_2 \ra
        \la Pb_1 n_2|I(\veps_{a_1}-\veps_{Pb_1})|a_1 a_2 \ra}
       {(\veps_{Pb_2}+k_{f_1}^0-\veps_{n_1})(E_A^{(0)}-\veps_{Pb_1}-\veps_{n_2})}\right.\nonumber\\
  &+&\left.\sum_{n_1,n_2}^{\veps_{Pb_2}+\veps_{n_2} \ne E_A^{(0)}}
  \frac{\la Pb_1|R_{f_1}|n_1 \ra \la n_1|R_{f_2}|n_2 \ra
        \la n_2 Pb_2|I(\veps_{a_2}-\veps_{Pb_2})|a_1 a_2 \ra}
       {(\veps_{Pb_1}+k_{f_1}^0-\veps_{n_1})(E_A^{(0)}-\veps_{Pb_2}-\veps_{n_2})}
  +(f_1 \leftrightarrow f_2)\right\}\,.
\ee
A similar calculation for the diagrams shown in Figs.~\ref{fig:2}(B)-\ref{fig:2}(D) yields
\be
\label{tau_B}
\tau_{\gamma_{f_1},\gamma_{f_2},B;A}^{(1B)} &=& -F_A F_B \sum_{P}(-1)^P
  \left\{\sum_{n_1,n_2}
  \frac{\la Pb_2|R_{f_1}|n_1 \ra \la Pb_1 n_1|I(\veps_{a_1}-\veps_{Pb_1})|a_1 n_2 \ra
        \la n_2|R_{f_2}|a_2 \ra}
       {(\veps_{Pb_2}+k_{f_1}^0-\veps_{n_1})(E_B^{(0)}-\veps_{a_1}+k_{f_1}^0-\veps_{n_2})}\right.\nonumber\\
  &+&\left.\sum_{n_1,n_2}
  \frac{\la Pb_1|R_{f_1}|n_1 \ra \la n_1 Pb_2|I(\veps_{a_2}-\veps_{Pb_2})|n_2 a_2 \ra
        \la n_2|R_{f_2}|a_1 \ra}
       {(\veps_{Pb_1}+k_{f_1}^0-\veps_{n_1})(E_B^{(0)}-\veps_{a_2}+k_{f_1}^0-\veps_{n_2})}
  +(f_1 \leftrightarrow f_2)\right\}\,,
\ee
\be \label{tau_C}
\tau_{\gamma_{f_1},\gamma_{f_2},B;A}^{(1C,{\rm irr})} &=& -F_A F_B \sum_{P}(-1)^P
  \left\{\sum_{n_1,n_2}^{\veps_{a_1}+\veps_{n_1} \ne E_B^{(0)}}
  \frac{\la Pb_1 Pb_2|I(\veps_{a_1}-\veps_{Pb_1})|a_1 n_1 \ra \la n_1|R_{f_1}|n_2 \ra
        \la n_2|R_{f_2}|a_2 \ra}
       {(E_B^{(0)}-\veps_{a_1}-\veps_{n_1})(E_B^{(0)}-\veps_{a_1}+k_{f_1}^0-\veps_{n_2})}\right.\nonumber\\
  &+&\left.\sum_{n_1,n_2}^{\veps_{a_2}+\veps_{n_1} \ne E_B^{(0)}}
  \frac{\la Pb_1 Pb_2|I(\veps_{a_2}-\veps_{Pb_2})|n_1 a_2 \ra \la n_1|R_{f_1}|n_2 \ra
        \la n_2|R_{f_2}|a_1 \ra}
       {(E_B^{(0)}-\veps_{a_2}-\veps_{n_1})(E_B^{(0)}-\veps_{a_2}+k_{f_1}^0-\veps_{n_2})}
  +(f_1 \leftrightarrow f_2)\right\}\,,
\ee
\be
\label{tau_D}
\tau_{\gamma_{f_1},\gamma_{f_2},B;A}^{(1D)} &=& -F_A F_B \sum_{P}(-1)^P
  \left\{  \sum_{n_1,n_2}
  \frac{\la Pb_1|R_{f_1}|n_1 \ra \la Pb_2|R_{f_2}|n_2 \ra
        \la n_1 n_2|I(\veps_{a_1}-\veps_{Pb_1}-k_{f_1}^0)|a_1 a_2 \ra}
       {(\veps_{Pb_1}+k_{f_1}^0-\veps_{n_1})(E_A^{(0)}-\veps_{Pb_1}-k_{f_1}^0-\veps_{n_2})}\right.\nonumber\\
  &+&      \sum_{n_1,n_2}
  \frac{\la Pb_1|R_{f_1}|n_1 \ra \la n_1 Pb_2|I(\veps_{a_1}-\veps_{Pb_1}-k_{f_1}^0)|a_1 n_2 \ra
        \la n_2|R_{f_2}|a_2 \ra}
       {(\veps_{Pb_1}+k_{f_1}^0-\veps_{n_1})(E_B^{(0)}-\veps_{a_1}+k_{f_1}^0-\veps_{n_2})}\nonumber\\
  &+&      \sum_{n_1,n_2}
  \frac{\la Pb_1 Pb_2|I(\veps_{a_1}-\veps_{Pb_1}-k_{f_1}^0)|n_1 n_2 \ra \la n_1|R_{f_1}|a_1 \ra 
        \la n_2|R_{f_2}|a_2 \ra}
       {(\veps_{a_1}-k_{f_1}^0-\veps_{n_1})(E_B^{(0)}-\veps_{a_1}+k_{f_1}^0-\veps_{n_2})}\nonumber\\
  &+&\left.\sum_{n_1,n_2}
  \frac{\la Pb_2|R_{f_2}|n_2 \ra
        \la Pb_1 n_2|I(\veps_{a_1}-\veps_{Pb_1}-k_{f_1}^0)|n_1 a_2 \ra \la n_1|R_{f_1}|a_1 \ra}
       {(\veps_{a_1}-k_{f_1}^0-\veps_{n_1})(E_A^{(0)}-\veps_{Pb_1}-k_{f_1}^0-\veps_{n_2})}
  +(f_1 \leftrightarrow f_2)\right\}\,.
\ee
In the case under consideration only the diagrams depicted in Figs.~\ref{fig:2}(A)
and \ref{fig:2}(C) possess reducible parts. For the reducible contribution coming
from the \ref{fig:2}(A) diagrams we have
\be
\label{tau_A_red}
\tau_{\gamma_{f_1},\gamma_{f_2},B;A}^{(1A,{\rm red})} &=& F_A F_B \sum_{P}(-1)^P \sum_n
  \left\{\frac{\la Pb_2|R_{f_2}|n\ra\la n|R_{f_1}|a_2\ra
           \delta_{Pb_1 a_1}}{(\veps_{a_2}-k_{f_1}^0-\veps_n)^2}
        +\frac{\la Pb_1|R_{f_2}|n\ra\la n|R_{f_1}|a_1\ra
           \delta_{Pb_2 a_2}}{(\veps_{a_1}-k_{f_1}^0-\veps_n)^2}\right\}
  \Delta E_A^{(1)}\nonumber\\
  &-&\frac{i}{2\pi}
  \tau_{\gamma_{f_1},\gamma_{f_2},B;A}^{(0)} F_{A'} F_{A''} \int_{-\infty}^{\infty} dp^0
  \left[\frac{\la a'_1 a'_2|I(p^0)|a''_1 a''_2 \ra}{(p^0+i0)^2}
       -\frac{\la a'_2 a'_1|I(p^0)|a''_1 a''_2 \ra}{(p^0-\Delta_A+i0)^2}\right]\,,
\ee
where $\Delta E_A^{(1)} =
F_{A'} F_{A''}\sum_{P}(-1)^P\la Pa'_1 Pa'_2|I(\veps_{Pa'_1}-\veps_{a''_1})|a''_1 a''_2 \ra$
is the one-photon exchange correction to the state $A$ and $\Delta_A = \veps_{a_2} - \veps_{a_1}$.
Combining this term together with the reducible part of the \ref{fig:2}(C) diagrams
and with the second term in formula (\ref{int1}), we obtain the total reducible contribution:
\be
\label{tau_red}
\tau_{\gamma_{f_1},\gamma_{f_2},B;A}^{(1,{\rm red})} &=& F_A F_B \sum_{P}(-1)^P \sum_n
  \left\{\frac{\la Pb_2|R_{f_1}|n\ra\la n|R_{f_2}|a_2\ra
           \delta_{Pb_1 a_1}}{(\veps_{Pb_2}+k_{f_1}^0-\veps_n)^2}\Delta E_B^{(1)}
        +\frac{\la Pb_1|R_{f_1}|n\ra\la n|R_{f_2}|a_1\ra
           \delta_{Pb_2 a_2}}{(\veps_{Pb_1}+k_{f_1}^0-\veps_n)^2}\Delta E_B^{(1)}\right.\nonumber\\
&+&\left.\frac{\la Pb_2|R_{f_2}|n\ra\la n|R_{f_1}|a_2\ra
           \delta_{Pb_1 a_1}}{(\veps_{a_2}-k_{f_1}^0-\veps_n)^2}\Delta E_A^{(1)}
        +\frac{\la Pb_1|R_{f_2}|n\ra\la n|R_{f_1}|a_1\ra
           \delta_{Pb_2 a_2}}{(\veps_{a_1}-k_{f_1}^0-\veps_n)^2}\Delta E_A^{(1)}\right\}\nonumber\\
  &+&\frac{1}{2}
  \tau_{\gamma_{f_1},\gamma_{f_2},B;A}^{(0)}
  \left[ F_{A'} F_{A''} \la a'_2 a'_1|I'(\Delta_A)|a''_1 a''_2 \ra +
    F_{B'} F_{B''} \la b'_2 b'_1|I'(\Delta_B)|b''_1 b''_2 \ra \right]\,,
\ee
where $\Delta E_B^{(1)}$ and $\Delta_B$ are defined similar as $\Delta E_A^{(1)}$ and $\Delta_A$,
$I'(\Delta) = [dI(\omega)/d\omega]_{\omega=\Delta}$.
The final expression for $\tau_{\gamma_{f_1},\gamma_{f_2},B;A}^{(1)}$
is given by the sum of Eqs.~(\ref{tau_A})-(\ref{tau_D}), and (\ref{tau_red}):
\be
\label{tau_1}
\tau_{\gamma_{f_1},\gamma_{f_2},B;A}^{(1)} =
   \tau_{\gamma_{f_1},\gamma_{f_2},B;A}^{(1A,{\rm irr})}
  +\tau_{\gamma_{f_1},\gamma_{f_2},B;A}^{(1B)}
  +\tau_{\gamma_{f_1},\gamma_{f_2},B;A}^{(1C,{\rm irr})}
  +\tau_{\gamma_{f_1},\gamma_{f_2},B;A}^{(1D)}
  +\tau_{\gamma_{f_1},\gamma_{f_2},B;A}^{(1,{\rm red})}\,.
\ee

Finally, the first-order interelectronic-interaction corrections to the
differential and total transition probabilities can be expressed according
to the following equations
\be
\label{dw_1}
dW_{B;A}^{(1)}(k_{f_1},\eps_{f_1},k_{f_2},\eps_{f_2}) &=&
  4\pi{\rm Re}\left\{\tau_{\gamma_{f_1},\gamma_{f_2},B;A}^{(0)*}
  \tau_{\gamma_{f_1},\gamma_{f_2},B;A}^{(1)}\right\}
  \delta(E_B^{(1)}+k_{f_1}^0+k_{f_2}^0-E_A^{(1)}) d\bfk_{f_1}d\bfk_{f_2}\nonumber\\
  &+& \Delta dW_{B;A}^{(0)}(k_{f_1},\eps_{f_1},k_{f_2},\eps_{f_2})\,,
\ee
\be
\label{w_1}
W_{B;A}^{(1)} = \frac{1}{2} \int_0^{\Delta^{(1)}_{AB}} dk_{f_1}^0\,
  (k_{f_1}^0)^2 (\Delta^{(1)}_{AB}-k_{f_1}^0)^2\,
  4\pi \sum_{\eps_{f_1},\eps_{f_2}} \int d\Omega_{k_{f_1}} d\Omega_{k_{f_2}}
  {\rm Re}\left\{\tau_{\gamma_{f_1},\gamma_{f_2},B;A}^{(0)*}
                  \tau_{\gamma_{f_1},\gamma_{f_2},B;A}^{(1)}\right\}
  + \Delta W_{B;A}^{(0)}\,,
\ee
where $E_A^{(1)} = E_A^{(0)} + \Delta E_A^{(1)}$, $E_B^{(1)} = E_B^{(0)}
+ \Delta E_B^{(1)}$, $\Delta^{(1)}_{AB} = E_A^{(1)} - E_B^{(1)}$, and
\be
\Delta dW_{B;A}^{(0)}(k_{f_1},\eps_{f_1},k_{f_2},\eps_{f_2}) =
  2\pi |\tau_{\gamma_{f_1},\gamma_{f_2},B;A}^{(0)}|^2
  \delta(E_B^{(1)}+k_{f_1}^0+k_{f_2}^0-E_A^{(1)}) d\bfk_{f_1}d\bfk_{f_2}
  - dW_{B;A}^{(0)}(k_{f_1},\eps_{f_1},k_{f_2},\eps_{f_2})\,,
\ee
\be
\Delta W_{B;A}^{(0)} = \frac{1}{2}\int_0^{\Delta^{(1)}_{AB}} dk_{f_1}^0\,
  (k_{f_1}^0)^2 (\Delta^{(1)}_{AB}-k_{f_1}^0)^2\,
  2\pi \sum_{\eps_{f_1},\eps_{f_2}} \int d\Omega_{k_{f_1}} d\Omega_{k_{f_2}}
  |\tau_{\gamma_{f_1},\gamma_{f_2},B;A}^{(0)}|^2 - W_{B;A}^{(0)}
\ee
are the contributions originating from changing the transition energy $\Delta^{(0)}_{AB}$
in the zeroth-order transition probability to the energy $\Delta^{(1)}_{AB}$,
which accounts for the interelectronic-interaction correction.
%
\subsection{First-order interelectronic-interaction correction with screening potential}
\label{sec:2C}

In the previous subsection we presented the formulas for the first-order
interelectronic-interaction correction involving electron states and propagators
in the external Coulomb potential of the nucleus as the zeroth-order approximation
(the original Furry picture). Now we consider an extended Furry picture, which
includes a local screening potential in the unperturbed Hamiltonian.
Since further we consider the two-photon decays from the single-excited state
to the ground state of He-like ions, we construct the screening potential for
the initial state $A$ such that it takes into account partly the
interelectronic interaction between the electrons $a_2$ and $a_1$. By employing the
extended Furry representation, we already at the zeroth-order level
relieve the quasidegeneracy of the $(1s2s)_J$ and $(1s2p_{1/2})_J$ states,
and improve the energy level scheme of the first
excited states in high-$Z$ heavy ions. Two different local screening potentials
$V_{\rm scr}$ are used: the Kohn-Sham potential and the core-Hartree potential.
Both potentials were successfully incorporated in previous calculations
\cite{glazov:2006:330,volotka:2008:062507,trotsenko:2010:033001,surzhykov:2010:042510}.

\begin{figure}
\includegraphics{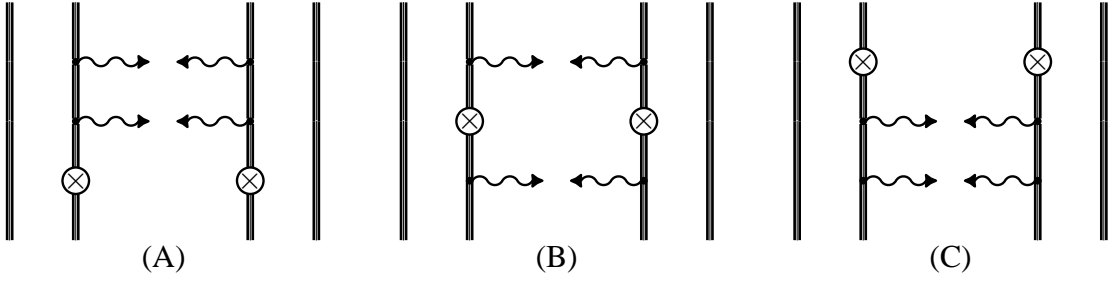}
\caption{The counterterm diagrams for the first-order interelectronic-interaction corrections
to the two-photon emission. The triple lines describe the electron propagators in the
effective potential. The symbol $\otimes$ represents the extra interaction term associated
with the local screening potential.
\label{fig:4}}
\end{figure}

\begin{figure}
\includegraphics{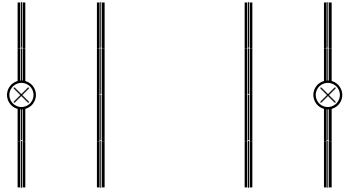}
\caption{The counterterm diagrams for the one-photon exchange correction.
Notations are the same as in Fig.~\ref{fig:4}.
\label{fig:5}}
\end{figure}

In the extended Furry picture we solve the Dirac equation with an effective
spherically symmetric potential treating the interaction with the external
Coulomb potential of the nucleus and the local screening potential exact to all
orders. The electron propagators in Figs.~\ref{fig:1}-\ref{fig:3}
have to be treated in the effective potential (we indicate this diagrammatically
via the triple electron line). The formulas derived in the previous subsection
remain formally the same, but keeping in mind that the Dirac spectrum
is now generated by solving the Dirac equation with the effective potential.
However, additional counterterm diagrams with the extra interaction term
$-V_{\rm scr}$ arise. In Figs.~\ref{fig:4} and \ref{fig:5} the additional diagrams are
depicted, where the extra interaction term $-V_{\rm scr}$ is represented graphically by
the symbol $\otimes$.
Thus, according to the Feynman rules we derive the expressions for the counterterm
diagrams shown in Figs.~\ref{fig:4}(A)-\ref{fig:4}(C)
\be
\label{tau_scr_A}
\tau_{\gamma_{f_1},\gamma_{f_2},B;A}^{(1A,{\rm irr}){\rm ext}} &=& F_A F_B \sum_{P}(-1)^P
  \left\{\sum_{n_1,n_2}^{\veps_{n_2} \ne \veps_{a_2}}
  \frac{\la Pb_2|R_{f_1}|n_1 \ra \la n_1|R_{f_2}|n_2 \ra
        \la n_2|V_{\rm scr}|a_2 \ra\delta_{Pb_1 a_1}}
       {(\veps_{Pb_2}+k_{f_1}^0-\veps_{n_1})(\veps_{a_2}-\veps_{n_2})}\right.\nonumber\\
  &+&\left.\sum_{n_1,n_2}^{\veps_{n_2} \ne \veps_{a_1}}
  \frac{\la Pb_1|R_{f_1}|n_1 \ra \la n_1|R_{f_2}|n_2 \ra
        \la n_2 |V_{\rm scr}|a_1 \ra\delta_{Pb_2 a_2}}
       {(\veps_{Pb_1}+k_{f_1}^0-\veps_{n_1})(\veps_{a_1}-\veps_{n_2})}
  +(f_1 \leftrightarrow f_2)\right\}\,,
\ee
\be
\label{tau_scr_B}
\tau_{\gamma_{f_1},\gamma_{f_2},B;A}^{(1B){\rm ext}} &=& F_A F_B \sum_{P}(-1)^P
  \left\{\sum_{n_1,n_2}
  \frac{\la Pb_2|R_{f_1}|n_1 \ra \la n_1|V_{\rm scr}|n_2 \ra
        \la n_2|R_{f_2}|a_2 \ra\delta_{Pb_1 a_1}}
       {(\veps_{Pb_2}+k_{f_1}^0-\veps_{n_1})(\veps_{Pb_2}+k_{f_1}^0-\veps_{n_2})}\right.\nonumber\\
  &+&\left.\sum_{n_1,n_2}
  \frac{\la Pb_1|R_{f_1}|n_1 \ra \la n_1|V_{\rm scr}|n_2 \ra
        \la n_2|R_{f_2}|a_1 \ra\delta_{Pb_2 a_2}}
       {(\veps_{Pb_1}+k_{f_1}^0-\veps_{n_1})(\veps_{Pb_1}+k_{f_1}^0-\veps_{n_2})}
  +(f_1 \leftrightarrow f_2)\right\}\,,
\ee
\be
\label{tau_scr_C}
\tau_{\gamma_{f_1},\gamma_{f_2},B;A}^{(1C,{\rm irr}){\rm ext}} &=& F_A F_B \sum_{P}(-1)^P
  \left\{\sum_{n_1,n_2}^{\veps_{n_1} \ne \veps_{Pb_2}}
  \frac{\la Pb_2|V_{\rm scr}|n_1 \ra \la n_1|R_{f_1}|n_2 \ra
        \la n_2|R_{f_2}|a_2 \ra\delta_{Pb_1 a_1}}
       {(\veps_{Pb_2}-\veps_{n_1})(\veps_{Pb_2}+k_{f_1}^0-\veps_{n_2})}\right.\nonumber\\
  &+&\left.\sum_{n_1,n_2}^{\veps_{n_1} \ne \veps_{Pb_1}}
  \frac{\la Pb_1|V_{\rm scr}|n_1 \ra \la n_1|R_{f_1}|n_2 \ra
        \la n_2|R_{f_2}|a_1 \ra\delta_{Pb_2 a_2}}
       {(\veps_{Pb_1}-\veps_{n_1})(\veps_{Pb_1}+k_{f_1}^0-\veps_{n_2})}
  +(f_1 \leftrightarrow f_2)\right\}\,.
\ee
For the additional reducible contribution we obtain
\be
\label{tau_scr_red}
\tau_{\gamma_{f_1},\gamma_{f_2},B;A}^{(1,{\rm red}){\rm ext}} &=& -F_A F_B \sum_{P}(-1)^P \sum_n
  \left\{\left[\frac{\la Pb_2|R_{f_1}|n\ra\la n|R_{f_2}|a_2\ra
     \delta_{Pb_1 a_1}}{(\veps_{Pb_2}+k_{f_1}^0-\veps_n)^2}
        +\frac{\la Pb_1|R_{f_1}|n\ra\la n|R_{f_2}|a_1\ra
     \delta_{Pb_2 a_2}}{(\veps_{Pb_1}+k_{f_1}^0-\veps_n)^2}\right]
     \Delta E_B^{(1){\rm ext}}\right.\nonumber\\
&+&\left.\left[\frac{\la Pb_2|R_{f_2}|n\ra\la n|R_{f_1}|a_2\ra
     \delta_{Pb_1 a_1}}{(\veps_{a_2}-k_{f_1}^0-\veps_n)^2}
        +\frac{\la Pb_1|R_{f_2}|n\ra\la n|R_{f_1}|a_1\ra
     \delta_{Pb_2 a_2}}{(\veps_{a_1}-k_{f_1}^0-\veps_n)^2}
     \right]\Delta E_A^{(1){\rm ext}}\right\}\,,
\ee
where $\Delta E_A^{(1){\rm ext}}$ and $\Delta E_B^{(1){\rm ext}}$ are the
counterterm contributions to the energy of the initial and final states,
respectively,
\be
\Delta E_A^{(1){\rm ext}} = -F_{A'} F_{A''}\sum_{P}(-1)^P (
   \la Pa'_1|V_{\rm scr}|a''_1 \ra\delta_{Pa'_2a''_2}
  +\la Pa'_2|V_{\rm scr}|a''_2 \ra\delta_{Pa'_1a''_1})\,.
\ee

Thus, in the extended Furry representation these extra terms have to be added
to the corresponding corrections to the transition amplitude as
$\tau_{\gamma_{f_1},\gamma_{f_2},B;A}^{(1A,{\rm irr})} \rightarrow
 \left(\tau_{\gamma_{f_1},\gamma_{f_2},B;A}^{(1A,{\rm irr})}
     + \tau_{\gamma_{f_1},\gamma_{f_2},B;A}^{(1A,{\rm irr}){\rm ext}}\right)$\,,
and, similarly, the rest terms.
Moreover, in Eqs.~(\ref{dw_1}) and (\ref{w_1}) the employed energies $E_A^{(1)}$
and $E_B^{(1)}$ have to be corrected to the counterterm contributions
$E_A^{(1)} = E_A^{(0)} + \Delta E_A^{(1)} + \Delta E_A^{(1){\rm ext}}$
and $E_B^{(1)} = E_B^{(0)} + \Delta E_B^{(1)} + \Delta E_B^{(1){\rm ext}}$.
%
\section{Numerical results and discussion}
\label{sec:3}

Now let us turn to the presentation and discussion of our numerical results for
the two-photon transitions $2^1S_0 \rightarrow 1^1S_0$ and $2^3S_1 \rightarrow 1^1S_0$
in He-like ions. The infinite summations over the complete Dirac spectrum
involved in the numerical evaluations are performed employing the finite-basis set method.
The B-splines basis set was constructed utilizing the dual kinetic balance approach
\cite{shabaev:2004:130405}. The homogeneously charged sphere model for the nuclear charge
distribution is employed together with the rms radii taken from Ref.~\cite{angeli:2004:185},
except for the thorium and uranium ions, for which the recent rms values are taken
from Ref.~\cite{kozhedub:2008:032501}. The Kohn-Sham and core-Hartree screening potentials
are employed in the zeroth-order approximation. The Kohn-Sham potentials are constructed
for the $2^1S_0$ state in the case of $2^1S_0 \rightarrow 1^1S_0$ transition, and for
the $2^3S_1$ state in the case of $2^3S_1 \rightarrow 1^1S_0$ transition, while
the core-Hartree potential is just a Coulomb potential generated by the $1s$ electron.
The screening potentials are generated self-consistently by solving the Dirac equation
until the energies of the core and valence states become stable on the level of $10^{-9}$.
In our final compilation we employ the Kohn-Sham potential as a starting one, since
the transition energies are better reproduced in this case.
The gauge invariance serves as an accurate check of consistency of the derived formulas
and the numerical procedure.  We analytically proof the gauge invariance of the obtained
formulas. In order to separate out the proper gauge invariant first-order
contribution we replace the transition operator $R_{f_2}$ with the first two terms
of the Taylor expansion in $\tau_{\gamma_{f_1},\gamma_{f_2},B;A}^{(0)}$, as
$R_{f_2}(k_{f_2}^0) \simeq R_{f_2}(\Delta_{AB}^{(0)}-k_{f_1}^0) +
 R'_{f_2}(\Delta_{AB}^{(0)}-k_{f_1}^0)\times\left[\Delta E_A^{(1)}-\Delta E_B^{(1)}\right]$,
and with the first term only in $\tau_{\gamma_{f_1},\gamma_{f_2},B;A}^{(1)}$, as
$R_{f_2}(k_{f_2}^0) \simeq R_{f_2}(\Delta_{AB}^{(0)}-k_{f_1}^0)$.
In the numerical procedure we employ the Feynman and Coulomb gauges for the
photon propagator and the velocity and length gauges for the emitted photons and
demonstrate the gauge independence of the final results.
In Table~\ref{tab:ind} we present the numerical results for the individual
contributions evaluated in the different gauges for He-like thorium.
As one can see from the table, the gauge invariance is restored in the final
values.
\begin{table}
\caption{Individual contributions to the total two-photon decay rates
for the transitions $2^1S_0 \rightarrow 1^1S_0 + 2\gamma(\rm E1)$
and $2^3S_1 \rightarrow 1^1S_0 + 2\gamma(\rm E1)$ in He-like $^{232}$Th$^{88+}$,
in units s$^{-1}$. The Kohn-Sham potential has been used as the starting potential.
The velocity and length gauges have been employed for the emitted photons,
and Feynman and Coulomb gauges for the photon propagator.
The more accurate transition energies $\Delta^{(1)}_{2^1S_0; 1^1S_0} = 91531$ eV and
$\Delta^{(1)}_{2^3S_1; 1^1S_0} = 91291$ eV are taken from Ref.~\cite{artemyev:2005:062104}.
Numbers in brackets are powers of ten.}
\label{tab:ind}
\tabcolsep5pt
\begin{tabular}{lccccc} \hline\hline
Gauges & $W_{B;A}^{(0)}$ & $\Delta W_{B;A}^{(0)}$ & $W_{B;A}^{(1,{\rm irr})}$
       & $W_{B;A}^{(1,{\rm red})}$ & $W_{B;A}$ \\ \hline
\multicolumn{6}{c}{$2^1S_0 \rightarrow 1^1S_0$}                                \\
Velocity / Feynman & 6.439[12]& -0.0862[12]& 0.0165[12]& 0.0123[12]& 6.381[12] \\
Length / Feynman   & 6.439[12]& -0.1610[12]& 0.0054[12]& 0.0982[12]& 6.381[12] \\
Velocity / Coulomb & 6.439[12]& -0.0862[12]& 0.0169[12]& 0.0119[12]& 6.381[12] \\
Length / Coulomb   & 6.439[12]& -0.1610[12]& 0.0058[12]& 0.0978[12]& 6.381[12] \\
\multicolumn{6}{c}{$2^3S_1 \rightarrow 1^1S_0$}                                \\
Velocity / Feynman & 1.686[10]& -0.0972[10]& 0.0115[10]& 0.0349[10]& 1.636[10] \\
Length / Feynman   & 1.686[10]& -0.1746[10]& 0.0369[10]& 0.0868[10]& 1.636[10] \\
Velocity / Coulomb & 1.686[10]& -0.0972[10]& 0.0114[10]& 0.0350[10]& 1.636[10] \\
Length / Coulomb   & 1.686[10]& -0.1746[10]& 0.0369[10]& 0.0869[10]& 1.636[10] \\
\hline\hline
\end{tabular}
\end{table}
A detailed discussion of these questions will be presented elsewhere.

In Table~\ref{tab:pot} we present the zeroth-order and final values of the
two-photon decay rates for the transitions $2^1S_0 \rightarrow 1^1S_0 + 2 \gamma ({\rm E1})$
and $2^3S_1 \rightarrow 1^1S_0 + 2 \gamma ({\rm E1})$ in He-like ions. 
These results include only the dominant 2E1 channel of the two-photon decay.
The final results for the total two-photon decay rates are evaluated according
to the following formula
\be
\label{w_tot}
W_{B;A} = \frac{1}{2} \int_0^{\Delta^{(1)}_{AB}} dk_{f_1}^0\,
  (k_{f_1}^0)^2 (\Delta^{(1)}_{AB}-k_{f_1}^0)^2\,
  2\pi \sum_{\eps_{f_1},\eps_{f_2}} \int d\Omega_{k_{f_1}} d\Omega_{k_{f_2}}
  |\tau_{\gamma_{f_1},\gamma_{f_2},B;A}^{(0)}
  +\tau_{\gamma_{f_1},\gamma_{f_2},B;A}^{(1)}|^2\,,
\ee
where in $\tau_{\gamma_{f_1},\gamma_{f_2},B;A}^{(0)}$ and
$\tau_{\gamma_{f_1},\gamma_{f_2},B;A}^{(1)}$, defined by Eq.~(\ref{tau_0})
and Eqs.~(\ref{tau_1}), (\ref{tau_scr_A})-(\ref{tau_scr_red}), respectively,
we separate out the terms up to the first order. The transition energies
$\Delta^{(1)}_{AB}$ together with the transition amplitudes
$\tau_{\gamma_{f_1},\gamma_{f_2},B;A}^{(1)}$ consistently
include the first-order interelectronic-interaction corrections to the two-photon decay
rate $W_{B;A}$. However, for high-$Z$ ions it is important also to
take into account the radiative corrections. In the framework of QED
perturbation theory, one has to evaluate radiative corrections
to both the transition energy and the transition amplitude.
In order to account partially for the radiative corrections,
we employ the more accurate transition energies taken from
Ref.~\cite{artemyev:2005:062104} for the transition energies $\Delta^{(1)}_{AB}$
in the upper integral limit and in the factor $(\Delta^{(1)}_{AB}-k_{f_1}^0)$
in Eq.~(\ref{w_tot}). Including by this way the more accurate transition
energies does not violate the gauge invariance of the result; it just scales
the decay rates to another value of the transition energy.
The employment of the more accurate transition energies yields corrections that
are negligible for intermediate-$Z$, which however become important for high-$Z$ ions.

The results of calculations performed by starting with the Coulomb, core-Hartree, and
Kohn-Sham potentials are presented in Table~\ref{tab:pot}.
Comparing the zeroth-order values in the Coulomb and screening potentials
one can observe that the screening potentials account for a considerable
part of electron-electron interaction effects. However, the difference
between the zeroth-order results for the core-Hartree and Kohn-Sham potentials 
is still quite large. Accounting for the first-order interelectronic-interaction
correction, we obtain the decay rates $W_{B;A}$, which much less depend on
the screening potential.
The remaining difference between the final values $W_{B;A}$ in the
core-Hartree and Kohn-Sham potentials provides a hint for the uncertainty
due to unaccounted second- and higher-order interelectronic-interaction corrections. 
In Table~\ref{tab:pot} we also compare the obtained decay rates $W_{B;A}$ with the
results of other theoretical calculations. In the case of the $2^1S_0$ state our
decay rates slightly disagree with the rates given by Derevianko and Johnson
\cite{derevianko:1997:1288}. For high-$Z$ ions this can
be explained by the radiative corrections, which are included in our
transition energies. The comparison with the results obtained by Drake
\cite{drake:1986:2871} gives a better agreement within the indicated uncertainty.  
In the case of the $2^3S_1$ state the interelectronic interaction
affects the two-photon decay rates much stronger, and therefore our
accuracy becomes slightly worse. For this case our results are in a fair
agreement with those values of Ref.~\cite{derevianko:1997:1288}.

As on can see from Table~\ref{tab:pot}, the final values of the total two-photon
decay rates calculated with the core-Hartree and Kohn-Sham potentials
are very close to each other. With this in mind, we restrict our further
consideration to the calculations performed with the Kohn-Sham screening
potential.
\begin{table}
\caption{The zeroth-order and final values of the total two-photon decay rates
(2E1 channel only) for the transitions $2^1S_0 \rightarrow 1^1S_0$ and
$2^3S_1 \rightarrow 1^1S_0$ in He-like ions starting with the Coulomb, core-Hartree,
and Kohn-Sham potentials, in units s$^{-1}$. Comparison with other theoretical
calculations is also made. Numbers in brackets denote powers of ten.}
\label{tab:pot}
\tabcolsep3mm
\begin{tabular}{llllllll} \hline\hline
    & Coulomb
      & \multicolumn{2}{c}{core-Hartree}
        & \multicolumn{2}{c}{Kohn-Sham}
          & \multicolumn{2}{c}{Other theor.}           \\
$Z$ & $W_{B;A}^{(0)}$
      & $W_{B;A}^{(0)}$ & $W_{B;A}$
        & $W_{B;A}^{(0)}$ & $W_{B;A}$
          & Ref.~\cite{derevianko:1997:1288}
            & Ref.~\cite{drake:1986:2871}              \\ \hline
\multicolumn{8}{c}{$2^1S_0 \rightarrow 1^1S_0$}        \\
 30 & 1.164[10] & 9.944[09] & 9.903[09]         
       & 1.006[10] & 9.900[09] & 9.938[09] & 9.88(3)[09] \\
 50 & 2.370[11] & 2.163[11] & 2.152[11]
       & 2.177[11] & 2.152[11] & 2.164[11] & 2.15(1)[11] \\
 70 & 1.655[12] & 1.554[12] & 1.545[12]
       & 1.560[12] & 1.544[12] & 1.556[12] & 1.55(1)[12] \\
 90 & 6.728[12] & 6.421[12] & 6.382[12]
       & 6.439[12] & 6.381[12] & 6.439[12] & 6.41(6)[12] \\
 92 & 7.580[12] & 7.242[12] & 7.199[12]
       & 7.262[12] & 7.199[12] & 7.265[12] & 7.24(8)[12] \\
\multicolumn{8}{c}{$2^3S_1 \rightarrow 1^1S_0$}          \\
 30 & 9.06[05] & 4.64[05] & 4.15[05]        
       & 4.42[05] & 4.13[05] & 4.17[05]  &             \\
 50 & 1.02[08] & 7.33[07] & 6.88[07]
       & 7.16[07] & 6.85[07] & 6.88[07]  &             \\
 70 & 2.13[09] & 1.74[09] & 1.67[09]
       & 1.72[09] & 1.66[09] & 1.66[09]  &             \\
 90 & 1.96[10] & 1.70[10] & 1.64[10]
       & 1.69[10] & 1.64[10] & 1.63[10]  &             \\
 92 & 2.38[10] & 2.07[10] & 1.99[10]
       & 2.05[10] & 1.99[10] & 1.98[10]  &             \\ \hline\hline
\end{tabular}
\end{table}

Beyond the dominant 2E1 decay channel we consider also the higher-multipole
contributions to the two-photon decay rates. In Table~\ref{tab:mp} we present
the contributions of higher multipoles calculated in the zeroth-order
approximation. In the case of the $2^1S_0$ state the contribution to the
total two-photon decay rate arises only from the photons with the same multipole
numbers. The correction due to the higher multipoles rapidly increases with $Z$,
but even for $Z = 92$ it is by a factor $10^3$ smaller than the dominant 2E1
decay rate. Unlike the $2^1S_0$ state, in the case of the two-photon
$2^3S_1 \rightarrow 1^1S_0$ transition the higher multipoles decay rates are
relatively large, as was first indicated by Dunford \cite{dunford:2004:062502}.
Our results for the E1M2 decay rate are in reasonable agreement with
those of Ref.~\cite{dunford:2004:062502}. Moreover, we also evaluate the 2M1
channel, which contribution becomes comparable with the E1M2 for high-$Z$ ions.
The contributions of higher multipoles are included in our final compilations.
\begin{table}
\caption{Contributions of the higher multipoles (MP) to the total two-photon
decay rates included in the zeroth-order approximation, in units s$^{-1}$.
Numbers in brackets denote powers of ten.}
\label{tab:mp}
\tabcolsep3mm
\begin{tabular}{llllll} \hline\hline
MP   &  $Z=30$  &  $Z=50$      &  $Z=70$      &  $Z=90$      &  $Z=92$      \\ \hline
\multicolumn{6}{c}{$2^1S_0 \rightarrow 1^1S_0$}                             \\
M1M1 & 1.40[04] & 2.65[06]     & 8.56[07]     & 1.14[09]     & 1.43[09]     \\
E2E2 & 4.74[03] & 8.19[05]     & 2.33[07]     & 2.71[08]     & 3.35[08]     \\
\multicolumn{6}{c}{$2^3S_1 \rightarrow 1^1S_0$}                             \\
E1M2 & 7.57[04] & 1.32[07]     & 3.79[08]     & 4.51[09]     & 5.59[09]     \\
     &          & 1.26[07]$^a$ & 3.62[08]$^a$ & 4.30[09]$^a$ & 5.32[09]$^a$ \\
M1M1 & 1.58[01] & 2.46[04]     & 3.33[06]     & 1.46[08]     & 2.05[08]     \\ \hline\hline
\end{tabular}
\newline
$^a$ Ref.~\cite{dunford:2004:062502}.
\end{table}

In Table~\ref{tab:exp} we compare the theoretical and experimental two-photon decay
rates of the $2^1S_0$ state. As one can see from the table, for Br$^{33+}$ and
Nb$^{39+}$ ions the theory is in a good agreement with the experiment, but for
Ni$^{26+}$ and Kr$^{34+}$ ions all theoretical calculations predict the values
being slightly larger than the experimental results. In the worst case of the Kr$^{34+}$
ion this difference amounts to about two standard deviations.
\begin{table}
\caption{Comparison of theory and experiment for the two-photon decay rates
of the $2^1S_0$ state in He-like ions, in units s$^{-1}$. Numbers in brackets
denote powers of ten.}
\label{tab:exp}
\tabcolsep3mm
\begin{tabular}{lllll} \hline\hline
$Z$ & Expt.             & This work & Ref.~\cite{derevianko:1997:1288}
                                                & Ref.~\cite{drake:1986:2871} \\ \hline
 28 & 6.406(66)[09]$^a$ & 6.493[09] & 6.517[09] & 6.482(21)[09]               \\
 35 & 2.543(21)[10]$^b$ & 2.529[10] & 2.540[10] &                             \\
 36 & 2.934(30)[10]$^c$ & 2.999[10] & 3.012[10] & 2.993(12)[10]               \\
 41 & 6.52(26)[10]$^d$  & 6.572[10] & 6.604[10] &                             \\ \hline\hline
\end{tabular}
\newline
$^a$ Ref.~\cite{dunford:1993:2729}.
\hspace{1cm}
$^c$ Ref.~\cite{marrus:1683:1986}.\\
$^b$ Ref.~\cite{dunford:1993:1929}.
\hspace{1cm}
$^d$ Ref.~\cite{simionovici:1695:1993}.
\end{table}
Finally, in Table~\ref{tab:total} we present our total two-photon decay rates
for the transitions $2^1S_0 \rightarrow 1^1S_0$ and $2^3S_1 \rightarrow 1^1S_0$.
\begin{table}
\caption{The total two-photon decay rates for the transitions
$2^1S_0 \rightarrow 1^1S_0$ and $2^3S_1 \rightarrow 1^1S_0$
in He-like ions, in units s$^{-1}$. The transition energies are
taken from Ref.~\cite{artemyev:2005:062104}.
Numbers in brackets denote powers of ten.}
\label{tab:total}
\tabcolsep3mm
\begin{tabular}{cccccc} \hline\hline
$Z$ & $2^1S_0$  & $2^3S_1$ & $Z$ & $2^1S_0$  & $2^3S_1$  \\ \hline
 28 & 6.493[09] & 2.40[05] &  61 & 6.948[11] & 5.56[08]  \\
 29 & 8.048[09] & 3.44[05] &  62 & 7.640[11] & 6.49[08]  \\
 30 & 9.900[09] & 4.88[05] &  63 & 8.388[11] & 7.56[08]  \\
 31 & 1.209[10] & 6.84[05] &  64 & 9.193[11] & 8.77[08]  \\
 32 & 1.467[10] & 9.46[05] &  65 & 1.006[12] & 1.02[09]  \\
 33 & 1.769[10] & 1.30[06] &  66 & 1.099[12] & 1.17[09]  \\
 34 & 2.121[10] & 1.76[06] &  67 & 1.199[12] & 1.35[09]  \\
 35 & 2.529[10] & 2.36[06] &  68 & 1.307[12] & 1.56[09]  \\
 36 & 2.999[10] & 3.13[06] &  69 & 1.422[12] & 1.79[09]  \\
 37 & 3.539[10] & 4.13[06] &  70 & 1.545[12] & 2.04[09]  \\
 38 & 4.158[10] & 5.40[06] &  71 & 1.676[12] & 2.33[09]  \\
 39 & 4.863[10] & 7.01[06] &  72 & 1.816[12] & 2.66[09]  \\
 40 & 5.664[10] & 9.04[06] &  73 & 1.966[12] & 3.03[09]  \\
 41 & 6.572[10] & 1.16[07] &  74 & 2.125[12] & 3.44[09]  \\
 42 & 7.595[10] & 1.47[07] &  75 & 2.294[12] & 3.90[09]  \\
 43 & 8.747[10] & 1.86[07] &  76 & 2.474[12] & 4.41[09]  \\
 44 & 1.004[11] & 2.33[07] &  77 & 2.665[12] & 4.98[09]  \\
 45 & 1.148[11] & 2.91[07] &  78 & 2.867[12] & 5.61[09]  \\
 46 & 1.310[11] & 3.61[07] &  79 & 3.082[12] & 6.32[09]  \\
 47 & 1.489[11] & 4.46[07] &  80 & 3.309[12] & 7.10[09]  \\
 48 & 1.688[11] & 5.49[07] &  81 & 3.549[12] & 7.96[09]  \\
 49 & 1.908[11] & 6.71[07] &  82 & 3.803[12] & 8.92[09]  \\
 50 & 2.152[11] & 8.18[07] &  83 & 4.071[12] & 9.98[09]  \\
 51 & 2.420[11] & 9.91[07] &  84 & 4.353[12] & 1.11[10]  \\
 52 & 2.715[11] & 1.20[08] &  85 & 4.650[12] & 1.24[10]  \\
 53 & 3.039[11] & 1.44[08] &  86 & 4.963[12] & 1.38[10]  \\
 54 & 3.394[11] & 1.73[08] &  87 & 5.292[12] & 1.54[10]  \\
 55 & 3.783[11] & 2.06[08] &  88 & 5.638[12] & 1.71[10]  \\
 56 & 4.206[11] & 2.45[08] &  89 & 6.002[12] & 1.90[10]  \\
 57 & 4.668[11] & 2.90[08] &  90 & 6.383[12] & 2.10[10]  \\
 58 & 5.171[11] & 3.43[08] &  91 & 6.782[12] & 2.32[10]  \\
 59 & 5.716[11] & 4.04[08] &  92 & 7.200[12] & 2.57[10]  \\
 60 & 6.308[11] & 4.75[08] &     &           &           \\ \hline\hline
\end{tabular}
\end{table}

\begin{figure}
\includegraphics[scale=0.7]{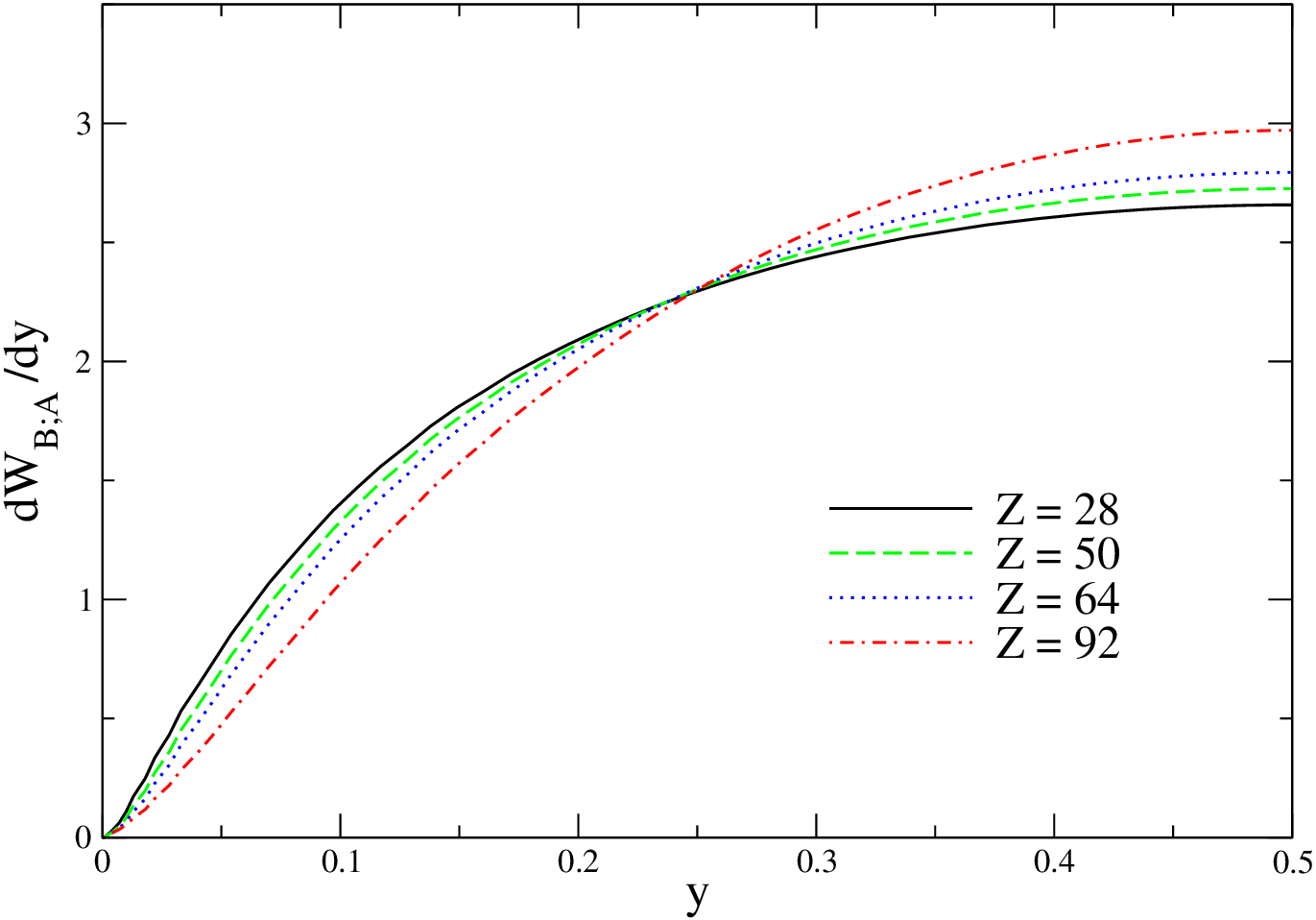}
\caption{(Color online) The $2^1S_0$ two-photon energy distribution functions
$dW_{B;A}/dy$, normalized to the corresponding total decay rates, plotted
as a function of the reduced energy $y$ for He-like nickel, tin, europium, and
uranium ions.
\label{fig:1s0}}
\end{figure}

\begin{figure}
\includegraphics[scale=0.7]{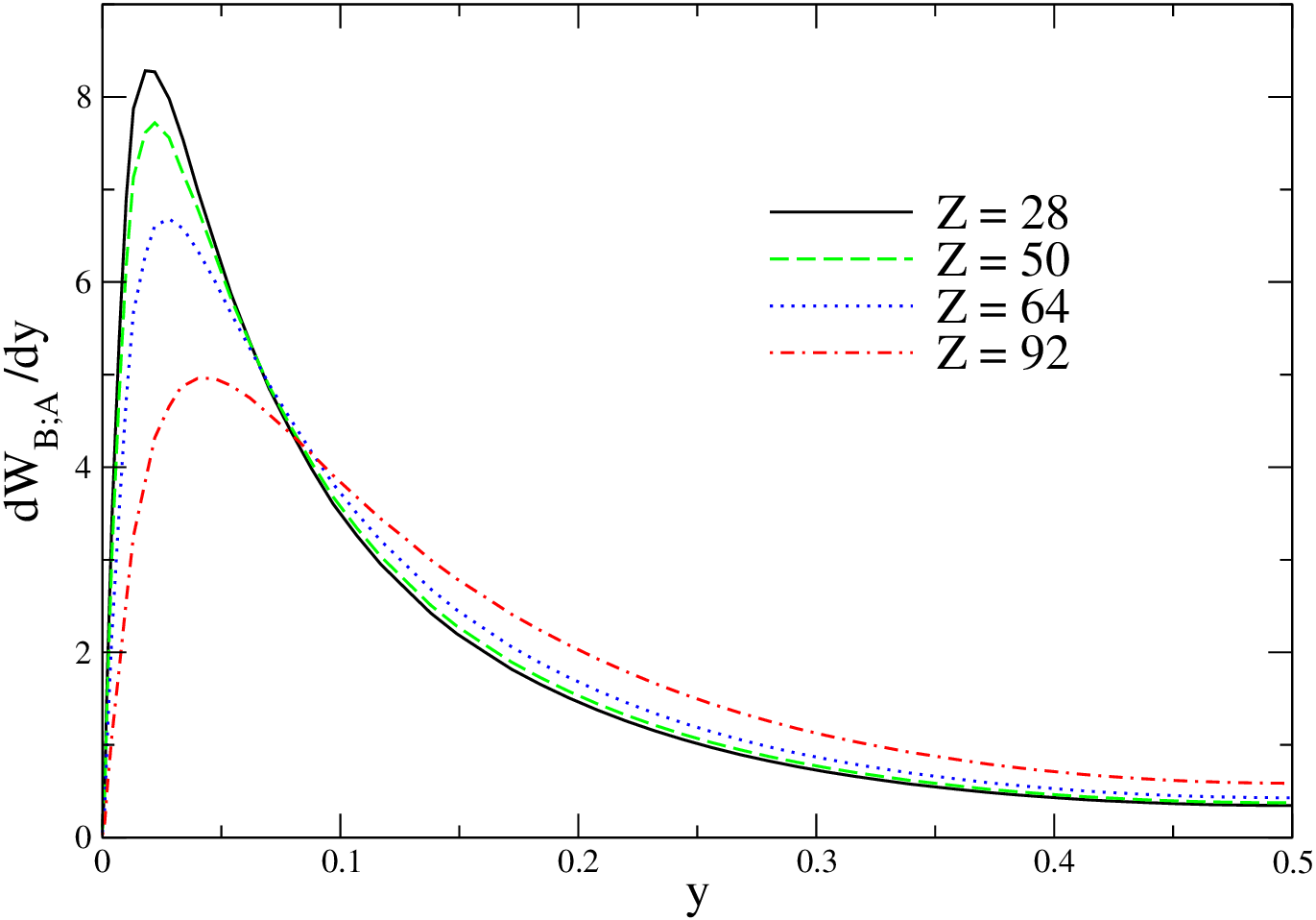}
\caption{(Color online) The $2^3S_1$ two-photon energy distribution functions
$dW_{B;A}/dy$, normalized to the corresponding total decay rates, plotted
as a function of the reduced energy $y$ for He-like nickel, tin, europium, and
uranium ions.
\label{fig:3s1}}
\end{figure}

Besides the total decay rates, we present the spectral-distribution functions
$dW_{B;A} / dy$ for the two-photon transitions $2^1S_0 \rightarrow 1^1S_0$
and $2^3S_1 \rightarrow 1^1S_0$ in Tables~\ref{tab:distr0} and \ref{tab:distr1},
respectively. The photon energy distribution function $dW_{B;A} / dy$
expressed as a function of the reduced energy $y = k_{f_1}^0 / \Delta^{(1)}_{AB}$
transported by one of the two photons reads
\be
\label{dw/dy}
dW_{B;A} / dy = y^2 (1-y)^2 (\Delta^{(1)}_{AB})^5\,
  2\pi \sum_{\eps_{f_1},\eps_{f_2}} \int d\Omega_{k_{f_1}} d\Omega_{k_{f_2}}
  |\tau_{\gamma_{f_1},\gamma_{f_2},B;A}^{(0)}
  +\tau_{\gamma_{f_1},\gamma_{f_2},B;A}^{(1)}|^2\,,
\ee
then the total decay rate can be found via the following equation
\be
W_{B;A} = \frac{1}{2} \int_0^1 dy\,\left(dW_{B;A} / dy\right)\,.
\ee
Since we employ the more accurate transition energy $\Delta^{(1)}_{AB}$ from
Ref.~\cite{artemyev:2005:062104}, our energy distribution function appears to be not
exactly symmetric with respect to the center point at $y = 0.5$. This asymmetry comes
mainly due to the higher-order corrections included in the transition energy but
neglected in the transition amplitude. In Tables~\ref{tab:distr0}, \ref{tab:distr1}
and in Figs.~\ref{fig:1s0}, \ref{fig:3s1} we present the spectral-distribution
functions $dW_{B;A} / dy$ calculated as a half-sum of the contributions
at the points $y$ and $1-y$. For the $2^1S_0$ state the energy distribution
function has one maximum at $y = 0.5$, and in Table~\ref{tab:distr0} we give also
the reduced full width at half maximum (FWHM) values. The behavior of the reduced
FWHM values as a function of $Z$ confirms those of Ref.~\cite{derevianko:1997:1288}.
For the $2^3S_1$ state the energy distribution function has two symmetric maxima
in first and second half of the unit segment.
In the center point (equal energy sharing) the distribution function is zero
for the decay channels with the photons with the same multipole numbers
(e.g., for the 2E1 decay). This is a consequence of the Bose-Einstein statistics,
which forbids to construct a permutation symmetric two-photon state with
total angular momentum $J_{\rm tot}=1$.
Therefore, near the center point the distribution function is defined by
the E1M2 channel, as it was noticed first in Ref.~\cite{dunford:2004:062502}.
The value of $y$, where the first
maximum is reached, is given in Table~\ref{tab:distr1} together with the
corresponding values of the reduced FWHM. In contrast to the results reported
in Ref.~\cite{derevianko:1997:1288}, we obtain a different energy distribution
due to accounting for the higher multipoles contributions. 
\begin{table}
\caption{The spectral distribution $dW_{B;A} / dy$ for the two-photon
transition $2^1S_0 \rightarrow 1^1S_0$ in He-like ions, in units s$^{-1}$.
The reduced photon energy $y = k_{f_1}^0 / \Delta^{(1)}_{AB}$ is the fraction
of the total transition energy transported by one of the two photons. The reduced
full widths at half maximum (FWHM) are also given. Numbers in brackets denote powers of ten.}
\label{tab:distr0}
\tabcolsep3mm
\begin{tabular}{cccccccccc} \hline\hline
 $y$  &  $Z=28$  &  $Z=36$  &  $Z=41$  &  $Z=50$  &  $Z=64$  &  $Z=70$  &  $Z=80$  &  $Z=90$  &  $Z=92$ \\ \hline
0.025 & 2.48[09] & 1.09[10] & 2.28[10] & 6.78[10] & 2.43[11] & 3.79[11] & 7.19[11] & 1.24[12] & 1.37[12]\\
0.050 & 5.16[09] & 2.31[10] & 4.92[10] & 1.52[11] & 5.77[11] & 9.17[11] & 1.78[12] & 3.11[12] & 3.44[12]\\
0.075 & 7.33[09] & 3.32[10] & 7.14[10] & 2.24[11] & 8.83[11] & 1.43[12] & 2.84[12] & 5.05[12] & 5.61[12]\\
0.100 & 9.10[09] & 4.14[10] & 8.97[10] & 2.85[11] & 1.15[12] & 1.88[12] & 3.80[12] & 6.90[12] & 7.68[12]\\
0.125 & 1.06[10] & 4.82[10] & 1.05[11] & 3.36[11] & 1.38[12] & 2.27[12] & 4.66[12] & 8.58[12] & 9.58[12]\\
0.150 & 1.18[10] & 5.39[10] & 1.17[11] & 3.79[11] & 1.57[12] & 2.61[12] & 5.42[12] & 1.01[13] & 1.13[13]\\
0.175 & 1.28[10] & 5.86[10] & 1.28[11] & 4.16[11] & 1.74[12] & 2.90[12] & 6.09[12] & 1.15[13] & 1.29[13]\\
0.200 & 1.36[10] & 6.26[10] & 1.37[11] & 4.47[11] & 1.89[12] & 3.15[12] & 6.67[12] & 1.27[13] & 1.42[13]\\
0.225 & 1.43[10] & 6.60[10] & 1.45[11] & 4.73[11] & 2.01[12] & 3.37[12] & 7.19[12] & 1.37[13] & 1.55[13]\\
0.250 & 1.49[10] & 6.89[10] & 1.51[11] & 4.96[11] & 2.12[12] & 3.57[12] & 7.64[12] & 1.47[13] & 1.66[13]\\
0.275 & 1.54[10] & 7.13[10] & 1.57[11] & 5.15[11] & 2.21[12] & 3.73[12] & 8.03[12] & 1.55[13] & 1.75[13]\\
0.300 & 1.58[10] & 7.34[10] & 1.61[11] & 5.31[11] & 2.29[12] & 3.87[12] & 8.37[12] & 1.63[13] & 1.84[13]\\
0.325 & 1.62[10] & 7.51[10] & 1.65[11] & 5.45[11] & 2.36[12] & 4.00[12] & 8.66[12] & 1.69[13] & 1.91[13]\\
0.350 & 1.65[10] & 7.66[10] & 1.68[11] & 5.57[11] & 2.42[12] & 4.10[12] & 8.90[12] & 1.74[13] & 1.97[13]\\
0.375 & 1.67[10] & 7.77[10] & 1.71[11] & 5.66[11] & 2.47[12] & 4.18[12] & 9.11[12] & 1.79[13] & 2.02[13]\\
0.400 & 1.69[10] & 7.87[10] & 1.73[11] & 5.74[11] & 2.50[12] & 4.25[12] & 9.27[12] & 1.82[13] & 2.07[13]\\
0.425 & 1.71[10] & 7.94[10] & 1.75[11] & 5.79[11] & 2.53[12] & 4.30[12] & 9.40[12] & 1.85[13] & 2.10[13]\\
0.450 & 1.72[10] & 7.99[10] & 1.76[11] & 5.83[11] & 2.55[12] & 4.34[12] & 9.49[12] & 1.87[13] & 2.12[13]\\
0.475 & 1.72[10] & 8.02[10] & 1.77[11] & 5.86[11] & 2.56[12] & 4.36[12] & 9.54[12] & 1.88[13] & 2.13[13]\\
0.500 & 1.73[10] & 8.03[10] & 1.77[11] & 5.87[11] & 2.57[12] & 4.37[12] & 9.56[12] & 1.89[13] & 2.14[13]\\ \hline
FWHM  & 0.814    & 0.809    & 0.804    & 0.793    & 0.771    & 0.761    & 0.743    & 0.722    & 0.718   \\ \hline\hline
\end{tabular}
\end{table}
\begin{table}
\caption{The spectral distribution $dW_{B;A} / dy$ for the two-photon
transition $2^3S_1 \rightarrow 1^1S_0$ in He-like ions, in units s$^{-1}$.
The reduced photon energy $y = k_{f_1}^0 / \Delta^{(1)}_{AB}$ is the fraction
of the total transition energy transported by one of the two photons. The
maximum point of the distribution $y_{\rm max}$ together with the reduced
full widths at half maximum (FWHM) are also presented. Numbers in brackets
denote powers of ten.}
\label{tab:distr1}
\tabcolsep3mm
\begin{tabular}{cccccccccc} \hline\hline
 $y$  &  $Z=28$  &  $Z=36$  &  $Z=41$  &  $Z=50$  &  $Z=64$  &  $Z=70$  &  $Z=80$  &  $Z=90$  &  $Z=92$ \\ \hline
0.010 & 1.68[06] & 2.27[07] & 8.13[07] & 5.12[08] & 4.22[09] & 8.62[09] & 2.40[10] & 5.71[10] & 6.70[10]\\
0.015 & 1.94[06] & 2.59[07] & 9.32[07] & 6.02[08] & 5.21[09] & 1.09[10] & 3.13[10] & 7.63[10] & 8.99[10]\\
0.020 & 1.99[06] & 2.64[07] & 9.56[07] & 6.30[08] & 5.68[09] & 1.21[10] & 3.57[10] & 8.92[10] & 1.05[11]\\
0.025 & 1.95[06] & 2.58[07] & 9.37[07] & 6.26[08] & 5.85[09] & 1.26[10] & 3.83[10] & 9.75[10] & 1.16[11]\\
0.030 & 1.87[06] & 2.46[07] & 8.98[07] & 6.08[08] & 5.83[09] & 1.28[10] & 3.95[10] & 1.03[11] & 1.22[11]\\
0.035 & 1.78[06] & 2.33[07] & 8.52[07] & 5.83[08] & 5.72[09] & 1.27[10] & 3.99[10] & 1.05[11] & 1.26[11]\\
0.040 & 1.68[06] & 2.19[07] & 8.04[07] & 5.55[08] & 5.55[09] & 1.24[10] & 3.97[10] & 1.06[11] & 1.27[11]\\
0.045 & 1.58[06] & 2.06[07] & 7.57[07] & 5.26[08] & 5.35[09] & 1.21[10] & 3.91[10] & 1.06[11] & 1.28[11]\\
0.050 & 1.48[06] & 1.94[07] & 7.12[07] & 4.98[08] & 5.14[09] & 1.17[10] & 3.83[10] & 1.05[11] & 1.27[11]\\
0.075 & 1.10[06] & 1.43[07] & 5.30[07] & 3.78[08] & 4.10[09] & 9.54[09] & 3.27[10] & 9.42[10] & 1.14[11]\\
0.100 & 8.40[05] & 1.09[07] & 4.05[07] & 2.93[08] & 3.27[09] & 7.72[09] & 2.72[10] & 8.08[10] & 9.87[10]\\
0.125 & 6.58[05] & 8.53[06] & 3.17[07] & 2.31[08] & 2.63[09] & 6.28[09] & 2.26[10] & 6.85[10] & 8.41[10]\\
0.150 & 5.25[05] & 6.81[06] & 2.54[07] & 1.86[08] & 2.14[09] & 5.15[09] & 1.88[10] & 5.80[10] & 7.15[10]\\
0.175 & 4.26[05] & 5.52[06] & 2.06[07] & 1.51[08] & 1.77[09] & 4.27[09] & 1.58[10] & 4.93[10] & 6.08[10]\\
0.200 & 3.50[05] & 4.54[06] & 1.69[07] & 1.25[08] & 1.47[09] & 3.57[09] & 1.33[10] & 4.20[10] & 5.20[10]\\
0.225 & 2.91[05] & 3.76[06] & 1.41[07] & 1.04[08] & 1.23[09] & 3.01[09] & 1.13[10] & 3.60[10] & 4.46[10]\\
0.250 & 2.43[05] & 3.15[06] & 1.18[07] & 8.75[07] & 1.04[09] & 2.55[09] & 9.63[09] & 3.09[10] & 3.84[10]\\
0.275 & 2.06[05] & 2.66[06] & 9.97[06] & 7.42[07] & 8.88[08] & 2.18[09] & 8.28[09] & 2.68[10] & 3.33[10]\\
0.300 & 1.75[05] & 2.27[06] & 8.51[06] & 6.34[07] & 7.63[08] & 1.88[09] & 7.16[09] & 2.33[10] & 2.90[10]\\
0.325 & 1.51[05] & 1.96[06] & 7.33[06] & 5.47[07] & 6.61[08] & 1.63[09] & 6.25[09] & 2.04[10] & 2.54[10]\\
0.350 & 1.31[05] & 1.70[06] & 6.38[06] & 4.77[07] & 5.79[08] & 1.43[09] & 5.50[09] & 1.80[10] & 2.25[10]\\
0.375 & 1.15[05] & 1.50[06] & 5.63[06] & 4.22[07] & 5.14[08] & 1.27[09] & 4.90[09] & 1.61[10] & 2.01[10]\\
0.400 & 1.03[05] & 1.34[06] & 5.05[06] & 3.79[07] & 4.62[08] & 1.15[09] & 4.43[09] & 1.46[10] & 1.83[10]\\
0.425 & 9.42[04] & 1.23[06] & 4.61[06] & 3.47[07] & 4.24[08] & 1.05[09] & 4.08[09] & 1.35[10] & 1.69[10]\\
0.450 & 8.80[04] & 1.15[06] & 4.31[06] & 3.24[07] & 3.97[08] & 9.87[08] & 3.83[09] & 1.27[10] & 1.59[10]\\
0.475 & 8.43[04] & 1.10[06] & 4.14[06] & 3.11[07] & 3.82[08] & 9.49[08] & 3.68[09] & 1.22[10] & 1.53[10]\\
0.500 & 8.31[04] & 1.08[06] & 4.08[06] & 3.07[07] & 3.77[08] & 9.36[08] & 3.64[09] & 1.21[10] & 1.51[10]\\ \hline
$y_{\rm max}$
      & 0.020    & 0.019    & 0.020    & 0.019    & 0.024    & 0.030    & 0.035    & 0.041    & 0.043   \\
FWHM  & 0.079    & 0.077    & 0.079    & 0.087    & 0.106    & 0.116    & 0.134    & 0.154    & 0.158   \\ \hline\hline
\end{tabular}
\end{table}
%
\section{Summary}

In summary, we have presented a systematic quantum electrodynamic description
for the first-order interelectronic-interaction corrections
to the two-photon transition probabilities in He-like ions.
A local screening potential has been included in the zeroth-order
approximation in the framework of an extended Furry representation,
and the corresponding expressions for the counterterms have been derived.
Such a treatment of the electron-correlation effects allows us to control
the gauge-invariance of each term in the perturbation expansion and to estimate
an uncertainty due to the truncation of this expansion.
The total two-photon decay rates and the spectral distribution functions
have been evaluated for the transitions $2^1S_0 \rightarrow 1^1S_0$ and
$2^3S_1 \rightarrow 1^1S_0$ in the He-like ions with nuclear charges in the
range $28 \le Z \le 92$. The results of the calculations performed have been
compared with previous calculations and with experimental data. The present calculations
of the two-photon decays of the $2^1S_0$ and $2^3S_1$ states in He-like ions
can be utilized for studying the parity non-conservation phenomena
in He-like ions \cite{schaefer:1989:7362,labzowsky:2001:054105,shabaev:2010:052102}
as well as for investigations of the contributions of higher multipoles
to the energy distribution.
%
\section*{Acknowledgments}
The authors owe thanks to D. A. Glazov and Th. St\"ohlker for valuable comments
and helpful discussions on this work. The work reported in this paper was supported
by the Helmholtz Gemeinschaft and GSI (Project VH-NG-421), by the Deutsche
Forschungsgemeinschaft (Grants No. VO 1707/1-1 and PL 254/7-1),
by RFBR (Grant No. 10-02-00450),
by the Russian Ministry of Education and Science (Grant No. P1334),
and by the grant of the President of the Russian Federation
(Grant No. MK-3215.2011.2).
Computing resources were provided by the Zentrum f\"ur Informationsdienste
und Hochleistungsrechnen (ZIH) at the TU Dresden.
%

%

\begin{thebibliography}{46}
\expandafter\ifx\csname natexlab\endcsname\relax\def\natexlab#1{#1}\fi
\expandafter\ifx\csname bibnamefont\endcsname\relax
  \def\bibnamefont#1{#1}\fi
\expandafter\ifx\csname bibfnamefont\endcsname\relax
  \def\bibfnamefont#1{#1}\fi
\expandafter\ifx\csname citenamefont\endcsname\relax
  \def\citenamefont#1{#1}\fi
\expandafter\ifx\csname url\endcsname\relax
  \def\url#1{\texttt{#1}}\fi
\expandafter\ifx\csname urlprefix\endcsname\relax\def\urlprefix{URL }\fi
\providecommand{\bibinfo}[2]{#2}
\providecommand{\eprint}[2][]{\url{#2}}

\bibitem[{\citenamefont{{G\"oppert-Mayer}}(1931)}]{goeppert:1931:273}
\bibinfo{author}{\bibfnamefont{M.}~\bibnamefont{{G\"oppert-Mayer}}},
  \bibinfo{journal}{Ann. Phys.} \textbf{\bibinfo{volume}{9}},
  \bibinfo{pages}{273} (\bibinfo{year}{1931}).

\bibitem[{\citenamefont{Drake}(1986)}]{drake:1986:2871}
\bibinfo{author}{\bibfnamefont{G.~W.~F.} \bibnamefont{Drake}},
  \bibinfo{journal}{Phys. Rev. A} \textbf{\bibinfo{volume}{34}},
  \bibinfo{pages}{2871} (\bibinfo{year}{1986}).

\bibitem[{\citenamefont{Santos et~al.}(1998)\citenamefont{Santos, Parente, and
  Indelicato}}]{santos:1998:43}
\bibinfo{author}{\bibfnamefont{J.~P.} \bibnamefont{Santos}},
  \bibinfo{author}{\bibfnamefont{F.}~\bibnamefont{Parente}}, \bibnamefont{and}
  \bibinfo{author}{\bibfnamefont{P.}~\bibnamefont{Indelicato}},
  \bibinfo{journal}{Eur. Phys. J. D} \textbf{\bibinfo{volume}{3}},
  \bibinfo{pages}{43} (\bibinfo{year}{1998}).

\bibitem[{\citenamefont{Surzhykov et~al.}(2005)\citenamefont{Surzhykov, Koval,
  and Fritzsche}}]{surzhykov:2005:022509}
\bibinfo{author}{\bibfnamefont{A.}~\bibnamefont{Surzhykov}},
  \bibinfo{author}{\bibfnamefont{P.}~\bibnamefont{Koval}}, \bibnamefont{and}
  \bibinfo{author}{\bibfnamefont{S.}~\bibnamefont{Fritzsche}},
  \bibinfo{journal}{Phys. Rev. A} \textbf{\bibinfo{volume}{71}},
  \bibinfo{pages}{022509} (\bibinfo{year}{2005}).

\bibitem[{\citenamefont{Labzowsky et~al.}(2005)\citenamefont{Labzowsky, Shonin,
  and Solovyev}}]{labzowsky:2005:265}
\bibinfo{author}{\bibfnamefont{L.~N.} \bibnamefont{Labzowsky}},
  \bibinfo{author}{\bibfnamefont{A.~V.} \bibnamefont{Shonin}},
  \bibnamefont{and} \bibinfo{author}{\bibfnamefont{D.~A.}
  \bibnamefont{Solovyev}}, \bibinfo{journal}{J. Phys. B}
  \textbf{\bibinfo{volume}{38}}, \bibinfo{pages}{265} (\bibinfo{year}{2005}).

\bibitem[{\citenamefont{Amaro et~al.}(2009)\citenamefont{Amaro, Santos,
  Parente, Surzhykov, and Indelicato}}]{amaro:2009:062504}
\bibinfo{author}{\bibfnamefont{P.}~\bibnamefont{Amaro}},
  \bibinfo{author}{\bibfnamefont{J.~P.} \bibnamefont{Santos}},
  \bibinfo{author}{\bibfnamefont{F.}~\bibnamefont{Parente}},
  \bibinfo{author}{\bibfnamefont{A.}~\bibnamefont{Surzhykov}},
  \bibnamefont{and}
  \bibinfo{author}{\bibfnamefont{P.}~\bibnamefont{Indelicato}},
  \bibinfo{journal}{Phys. Rev. A} \textbf{\bibinfo{volume}{79}},
  \bibinfo{pages}{062504} (\bibinfo{year}{2009}).

\bibitem[{\citenamefont{Dalgarno}(1966)}]{dalgarno:1966:311}
\bibinfo{author}{\bibfnamefont{A.}~\bibnamefont{Dalgarno}},
  \bibinfo{journal}{Mon. Not. R. Astron. Soc.} \textbf{\bibinfo{volume}{131}},
  \bibinfo{pages}{311} (\bibinfo{year}{1966}).

\bibitem[{\citenamefont{Bely and Faucher}(1969)}]{bely:1969:37}
\bibinfo{author}{\bibfnamefont{O.}~\bibnamefont{Bely}} \bibnamefont{and}
  \bibinfo{author}{\bibfnamefont{P.}~\bibnamefont{Faucher}},
  \bibinfo{journal}{Astron. Astrophys.} \textbf{\bibinfo{volume}{1}},
  \bibinfo{pages}{37} (\bibinfo{year}{1969}).

\bibitem[{\citenamefont{Drake et~al.}(1969)\citenamefont{Drake, Victor, and
  Dalgarno}}]{drake:1969:25}
\bibinfo{author}{\bibfnamefont{G.~W.~F.} \bibnamefont{Drake}},
  \bibinfo{author}{\bibfnamefont{G.~A.} \bibnamefont{Victor}},
  \bibnamefont{and} \bibinfo{author}{\bibfnamefont{A.}~\bibnamefont{Dalgarno}},
  \bibinfo{journal}{Phys. Rev.} \textbf{\bibinfo{volume}{180}},
  \bibinfo{pages}{25} (\bibinfo{year}{1969}).

\bibitem[{\citenamefont{Derevianko and Johnson}(1997)}]{derevianko:1997:1288}
\bibinfo{author}{\bibfnamefont{A.}~\bibnamefont{Derevianko}} \bibnamefont{and}
  \bibinfo{author}{\bibfnamefont{W.~R.} \bibnamefont{Johnson}},
  \bibinfo{journal}{Phys. Rev. A} \textbf{\bibinfo{volume}{56}},
  \bibinfo{pages}{1288} (\bibinfo{year}{1997}).

\bibitem[{\citenamefont{Surzhykov et~al.}(2010)\citenamefont{Surzhykov,
  Volotka, Fratini, Santos, Indelicato, Plunien, {Th.~St\"ohlker}, and
  Fritzsche}}]{surzhykov:2010:042510}
\bibinfo{author}{\bibfnamefont{A.}~\bibnamefont{Surzhykov}},
  \bibinfo{author}{\bibfnamefont{A.}~\bibnamefont{Volotka}},
  \bibinfo{author}{\bibfnamefont{F.}~\bibnamefont{Fratini}},
  \bibinfo{author}{\bibfnamefont{J.~P.} \bibnamefont{Santos}},
  \bibinfo{author}{\bibfnamefont{P.}~\bibnamefont{Indelicato}},
  \bibinfo{author}{\bibfnamefont{G.}~\bibnamefont{Plunien}},
  \bibinfo{author}{\bibnamefont{{Th.~St\"ohlker}}}, \bibnamefont{and}
  \bibinfo{author}{\bibfnamefont{S.}~\bibnamefont{Fritzsche}},
  \bibinfo{journal}{Phys. Rev. A} \textbf{\bibinfo{volume}{81}},
  \bibinfo{pages}{042510} (\bibinfo{year}{2010}).

\bibitem[{\citenamefont{Marrus et~al.}(1986)\citenamefont{Marrus, San~Vicente,
  Charles, Briand, Bosch, Liesen, and Varga}}]{marrus:1683:1986}
\bibinfo{author}{\bibfnamefont{R.}~\bibnamefont{Marrus}},
  \bibinfo{author}{\bibfnamefont{V.}~\bibnamefont{San~Vicente}},
  \bibinfo{author}{\bibfnamefont{P.}~\bibnamefont{Charles}},
  \bibinfo{author}{\bibfnamefont{J.~P.} \bibnamefont{Briand}},
  \bibinfo{author}{\bibfnamefont{F.}~\bibnamefont{Bosch}},
  \bibinfo{author}{\bibfnamefont{D.}~\bibnamefont{Liesen}}, \bibnamefont{and}
  \bibinfo{author}{\bibfnamefont{I.}~\bibnamefont{Varga}},
  \bibinfo{journal}{Phys. Rev. Lett.} \textbf{\bibinfo{volume}{56}},
  \bibinfo{pages}{1683} (\bibinfo{year}{1986}).

\bibitem[{\citenamefont{Dunford
  et~al.}(1993{\natexlab{a}})\citenamefont{Dunford, Berry, Cheng, Kanter,
  Kurtz, Zabransky, Livingston, and Curtis}}]{dunford:1993:1929}
\bibinfo{author}{\bibfnamefont{R.~W.} \bibnamefont{Dunford}},
  \bibinfo{author}{\bibfnamefont{H.~G.} \bibnamefont{Berry}},
  \bibinfo{author}{\bibfnamefont{S.}~\bibnamefont{Cheng}},
  \bibinfo{author}{\bibfnamefont{E.~P.} \bibnamefont{Kanter}},
  \bibinfo{author}{\bibfnamefont{C.}~\bibnamefont{Kurtz}},
  \bibinfo{author}{\bibfnamefont{B.~J.} \bibnamefont{Zabransky}},
  \bibinfo{author}{\bibfnamefont{A.~E.} \bibnamefont{Livingston}},
  \bibnamefont{and} \bibinfo{author}{\bibfnamefont{L.~J.}
  \bibnamefont{Curtis}}, \bibinfo{journal}{Phys. Rev. A}
  \textbf{\bibinfo{volume}{48}}, \bibinfo{pages}{1929}
  (\bibinfo{year}{1993}{\natexlab{a}}).

\bibitem[{\citenamefont{Dunford
  et~al.}(1993{\natexlab{b}})\citenamefont{Dunford, Berry, Church, Hass, Liu,
  Raphaelian, Zabransky, Curtis, and Livingston}}]{dunford:1993:2729}
\bibinfo{author}{\bibfnamefont{R.~W.} \bibnamefont{Dunford}},
  \bibinfo{author}{\bibfnamefont{H.~G.} \bibnamefont{Berry}},
  \bibinfo{author}{\bibfnamefont{D.~A.} \bibnamefont{Church}},
  \bibinfo{author}{\bibfnamefont{M.}~\bibnamefont{Hass}},
  \bibinfo{author}{\bibfnamefont{C.~J.} \bibnamefont{Liu}},
  \bibinfo{author}{\bibfnamefont{M.~L.~A.} \bibnamefont{Raphaelian}},
  \bibinfo{author}{\bibfnamefont{B.~J.} \bibnamefont{Zabransky}},
  \bibinfo{author}{\bibfnamefont{L.~J.} \bibnamefont{Curtis}},
  \bibnamefont{and} \bibinfo{author}{\bibfnamefont{A.~E.}
  \bibnamefont{Livingston}}, \bibinfo{journal}{Phys. Rev. A}
  \textbf{\bibinfo{volume}{48}}, \bibinfo{pages}{2729}
  (\bibinfo{year}{1993}{\natexlab{b}}).

\bibitem[{\citenamefont{Mokler et~al.}(1990)\citenamefont{Mokler, Reusch,
  Warczak, Stachura, Kambara, M{\"u}ller, Schuch, and
  Schulz}}]{mokler:1990:3108}
\bibinfo{author}{\bibfnamefont{P.~H.} \bibnamefont{Mokler}},
  \bibinfo{author}{\bibfnamefont{S.}~\bibnamefont{Reusch}},
  \bibinfo{author}{\bibfnamefont{A.}~\bibnamefont{Warczak}},
  \bibinfo{author}{\bibfnamefont{Z.}~\bibnamefont{Stachura}},
  \bibinfo{author}{\bibfnamefont{T.}~\bibnamefont{Kambara}},
  \bibinfo{author}{\bibfnamefont{A.}~\bibnamefont{M{\"u}ller}},
  \bibinfo{author}{\bibfnamefont{R.}~\bibnamefont{Schuch}}, \bibnamefont{and}
  \bibinfo{author}{\bibfnamefont{M.}~\bibnamefont{Schulz}},
  \bibinfo{journal}{Phys. Rev. Lett.} \textbf{\bibinfo{volume}{65}},
  \bibinfo{pages}{3108} (\bibinfo{year}{1990}).

\bibitem[{\citenamefont{Ali et~al.}(1997)\citenamefont{Ali, Ahmad, Dunford,
  Gemmell, Jung, Kanter, Mokler, Berry, Livingston, Cheng
  et~al.}}]{ali:1997:994}
\bibinfo{author}{\bibfnamefont{R.}~\bibnamefont{Ali}},
  \bibinfo{author}{\bibfnamefont{I.}~\bibnamefont{Ahmad}},
  \bibinfo{author}{\bibfnamefont{R.~W.} \bibnamefont{Dunford}},
  \bibinfo{author}{\bibfnamefont{D.~S.} \bibnamefont{Gemmell}},
  \bibinfo{author}{\bibfnamefont{M.}~\bibnamefont{Jung}},
  \bibinfo{author}{\bibfnamefont{E.~P.} \bibnamefont{Kanter}},
  \bibinfo{author}{\bibfnamefont{P.~H.} \bibnamefont{Mokler}},
  \bibinfo{author}{\bibfnamefont{H.~G.} \bibnamefont{Berry}},
  \bibinfo{author}{\bibfnamefont{A.~E.} \bibnamefont{Livingston}},
  \bibinfo{author}{\bibfnamefont{S.}~\bibnamefont{Cheng}},
  \bibnamefont{et~al.}, \bibinfo{journal}{Phys. Rev. A}
  \textbf{\bibinfo{volume}{55}}, \bibinfo{pages}{994} (\bibinfo{year}{1997}).

\bibitem[{\citenamefont{Sch{\"a}ffer et~al.}(1999)\citenamefont{Sch{\"a}ffer,
  Mokler, Dunford, Kozhuharov, Kr{\"a}mer, Livingston, Ludziejewski, Prinz,
  Rymuza, Sarkadi et~al.}}]{schaeffer:1999:489}
\bibinfo{author}{\bibfnamefont{H.~W.} \bibnamefont{Sch{\"a}ffer}},
  \bibinfo{author}{\bibfnamefont{P.~H.} \bibnamefont{Mokler}},
  \bibinfo{author}{\bibfnamefont{R.~W.} \bibnamefont{Dunford}},
  \bibinfo{author}{\bibfnamefont{C.}~\bibnamefont{Kozhuharov}},
  \bibinfo{author}{\bibfnamefont{A.}~\bibnamefont{Kr{\"a}mer}},
  \bibinfo{author}{\bibfnamefont{A.~E.} \bibnamefont{Livingston}},
  \bibinfo{author}{\bibfnamefont{T.}~\bibnamefont{Ludziejewski}},
  \bibinfo{author}{\bibfnamefont{H.-T.} \bibnamefont{Prinz}},
  \bibinfo{author}{\bibfnamefont{P.}~\bibnamefont{Rymuza}},
  \bibinfo{author}{\bibfnamefont{L.}~\bibnamefont{Sarkadi}},
  \bibnamefont{et~al.}, \bibinfo{journal}{Phys. Lett. A}
  \textbf{\bibinfo{volume}{260}}, \bibinfo{pages}{489} (\bibinfo{year}{1999}).

\bibitem[{\citenamefont{Kumar et~al.}(2009)\citenamefont{Kumar, Trotsenko,
  Volotka, Bana{\'s}, Beyer, Br{\"a}uning, Gumberidze, Hagmann, Hess,
  Kozhuharov et~al.}}]{kumar:2009:19}
\bibinfo{author}{\bibfnamefont{A.}~\bibnamefont{Kumar}},
  \bibinfo{author}{\bibfnamefont{S.}~\bibnamefont{Trotsenko}},
  \bibinfo{author}{\bibfnamefont{A.~V.} \bibnamefont{Volotka}},
  \bibinfo{author}{\bibfnamefont{D.}~\bibnamefont{Bana{\'s}}},
  \bibinfo{author}{\bibfnamefont{H.~F.} \bibnamefont{Beyer}},
  \bibinfo{author}{\bibfnamefont{H.}~\bibnamefont{Br{\"a}uning}},
  \bibinfo{author}{\bibfnamefont{A.}~\bibnamefont{Gumberidze}},
  \bibinfo{author}{\bibfnamefont{S.}~\bibnamefont{Hagmann}},
  \bibinfo{author}{\bibfnamefont{S.}~\bibnamefont{Hess}},
  \bibinfo{author}{\bibfnamefont{C.}~\bibnamefont{Kozhuharov}},
  \bibnamefont{et~al.}, \bibinfo{journal}{Eur. Phys. J. Special Topics}
  \textbf{\bibinfo{volume}{169}}, \bibinfo{pages}{19} (\bibinfo{year}{2009}).

\bibitem[{\citenamefont{Trotsenko et~al.}(2010)\citenamefont{Trotsenko, Kumar,
  Volotka, Bana\'s, Beyer, Br\"auning, Fritzsche, Gumberidze, Hagmann, Hess
  et~al.}}]{trotsenko:2010:033001}
\bibinfo{author}{\bibfnamefont{S.}~\bibnamefont{Trotsenko}},
  \bibinfo{author}{\bibfnamefont{A.}~\bibnamefont{Kumar}},
  \bibinfo{author}{\bibfnamefont{A.~V.} \bibnamefont{Volotka}},
  \bibinfo{author}{\bibfnamefont{D.}~\bibnamefont{Bana\'s}},
  \bibinfo{author}{\bibfnamefont{H.~F.} \bibnamefont{Beyer}},
  \bibinfo{author}{\bibfnamefont{H.}~\bibnamefont{Br\"auning}},
  \bibinfo{author}{\bibfnamefont{S.}~\bibnamefont{Fritzsche}},
  \bibinfo{author}{\bibfnamefont{A.}~\bibnamefont{Gumberidze}},
  \bibinfo{author}{\bibfnamefont{S.}~\bibnamefont{Hagmann}},
  \bibinfo{author}{\bibfnamefont{S.}~\bibnamefont{Hess}}, \bibnamefont{et~al.},
  \bibinfo{journal}{Phys. Rev. Lett.} \textbf{\bibinfo{volume}{104}},
  \bibinfo{pages}{033001} (\bibinfo{year}{2010}).

\bibitem[{\citenamefont{Savukov and Johnson}(2002)}]{savukov:2002:062507}
\bibinfo{author}{\bibfnamefont{I.~M.} \bibnamefont{Savukov}} \bibnamefont{and}
  \bibinfo{author}{\bibfnamefont{W.~R.} \bibnamefont{Johnson}},
  \bibinfo{journal}{Phys. Rev. A} \textbf{\bibinfo{volume}{66}},
  \bibinfo{pages}{062507} (\bibinfo{year}{2002}).

\bibitem[{\citenamefont{Shabaev et~al.}(2010)\citenamefont{Shabaev, Volotka,
  Kozhuharov, Plunien, and {Th.~St\"ohlker}}}]{shabaev:2010:052102}
\bibinfo{author}{\bibfnamefont{V.~M.} \bibnamefont{Shabaev}},
  \bibinfo{author}{\bibfnamefont{A.~V.} \bibnamefont{Volotka}},
  \bibinfo{author}{\bibfnamefont{C.}~\bibnamefont{Kozhuharov}},
  \bibinfo{author}{\bibfnamefont{G.}~\bibnamefont{Plunien}}, \bibnamefont{and}
  \bibinfo{author}{\bibnamefont{{Th.~St\"ohlker}}}, \bibinfo{journal}{Phys.
  Rev. A} \textbf{\bibinfo{volume}{81}}, \bibinfo{pages}{052102}
  (\bibinfo{year}{2010}).

\bibitem[{\citenamefont{Lapierre et~al.}(2005)\citenamefont{Lapierre,
  Jentschura, {Crespo~L\'opez-Urrutia}, Braun, Brenner, Bruhns, Fischer,
  {Gonz\'alez~Mart\'inez}, Harman, Johnson et~al.}}]{lapierre:2005:183001}
\bibinfo{author}{\bibfnamefont{A.}~\bibnamefont{Lapierre}},
  \bibinfo{author}{\bibfnamefont{U.~D.} \bibnamefont{Jentschura}},
  \bibinfo{author}{\bibfnamefont{J.~R.}
  \bibnamefont{{Crespo~L\'opez-Urrutia}}},
  \bibinfo{author}{\bibfnamefont{J.}~\bibnamefont{Braun}},
  \bibinfo{author}{\bibfnamefont{G.}~\bibnamefont{Brenner}},
  \bibinfo{author}{\bibfnamefont{H.}~\bibnamefont{Bruhns}},
  \bibinfo{author}{\bibfnamefont{D.}~\bibnamefont{Fischer}},
  \bibinfo{author}{\bibfnamefont{A.~J.} \bibnamefont{{Gonz\'alez~Mart\'inez}}},
  \bibinfo{author}{\bibfnamefont{Z.}~\bibnamefont{Harman}},
  \bibinfo{author}{\bibfnamefont{W.~R.} \bibnamefont{Johnson}},
  \bibnamefont{et~al.}, \bibinfo{journal}{Phys. Rev. Lett.}
  \textbf{\bibinfo{volume}{95}}, \bibinfo{pages}{183001}
  (\bibinfo{year}{2005}).

\bibitem[{\citenamefont{Lapierre et~al.}(2006)\citenamefont{Lapierre,
  {Crespo~L\'opez-Urrutia}, Braun, Brenner, Bruhns, Fischer,
  {Gonz\'alez~Mart\'inez}, Mironov, Osborne, Sikler
  et~al.}}]{lapierre:2006:052507}
\bibinfo{author}{\bibfnamefont{A.}~\bibnamefont{Lapierre}},
  \bibinfo{author}{\bibfnamefont{J.~R.}
  \bibnamefont{{Crespo~L\'opez-Urrutia}}},
  \bibinfo{author}{\bibfnamefont{J.}~\bibnamefont{Braun}},
  \bibinfo{author}{\bibfnamefont{G.}~\bibnamefont{Brenner}},
  \bibinfo{author}{\bibfnamefont{H.}~\bibnamefont{Bruhns}},
  \bibinfo{author}{\bibfnamefont{D.}~\bibnamefont{Fischer}},
  \bibinfo{author}{\bibfnamefont{A.~J.} \bibnamefont{{Gonz\'alez~Mart\'inez}}},
  \bibinfo{author}{\bibfnamefont{V.}~\bibnamefont{Mironov}},
  \bibinfo{author}{\bibfnamefont{C.}~\bibnamefont{Osborne}},
  \bibinfo{author}{\bibfnamefont{G.}~\bibnamefont{Sikler}},
  \bibnamefont{et~al.}, \bibinfo{journal}{Phys. Rev. A}
  \textbf{\bibinfo{volume}{73}}, \bibinfo{pages}{052507}
  (\bibinfo{year}{2006}).

\bibitem[{\citenamefont{Tupitsyn et~al.}(2005)\citenamefont{Tupitsyn, Volotka,
  Glazov, Shabaev, Plunien, {Crespo~L\'opez-Urrutia}, Lapierre, and
  Ullrich}}]{tupitsyn:2005:062503}
\bibinfo{author}{\bibfnamefont{I.~I.} \bibnamefont{Tupitsyn}},
  \bibinfo{author}{\bibfnamefont{A.~V.} \bibnamefont{Volotka}},
  \bibinfo{author}{\bibfnamefont{D.~A.} \bibnamefont{Glazov}},
  \bibinfo{author}{\bibfnamefont{V.~M.} \bibnamefont{Shabaev}},
  \bibinfo{author}{\bibfnamefont{G.}~\bibnamefont{Plunien}},
  \bibinfo{author}{\bibfnamefont{J.~R.}
  \bibnamefont{{Crespo~L\'opez-Urrutia}}},
  \bibinfo{author}{\bibfnamefont{A.}~\bibnamefont{Lapierre}}, \bibnamefont{and}
  \bibinfo{author}{\bibfnamefont{J.}~\bibnamefont{Ullrich}},
  \bibinfo{journal}{Phys. Rev. A} \textbf{\bibinfo{volume}{72}},
  \bibinfo{pages}{062503} (\bibinfo{year}{2005}).

\bibitem[{\citenamefont{Volotka et~al.}(2006)\citenamefont{Volotka, Glazov,
  Plunien, Shabaev, and Tupitsyn}}]{volotka:2006:293}
\bibinfo{author}{\bibfnamefont{A.~V.} \bibnamefont{Volotka}},
  \bibinfo{author}{\bibfnamefont{D.~A.} \bibnamefont{Glazov}},
  \bibinfo{author}{\bibfnamefont{G.}~\bibnamefont{Plunien}},
  \bibinfo{author}{\bibfnamefont{V.~M.} \bibnamefont{Shabaev}},
  \bibnamefont{and} \bibinfo{author}{\bibfnamefont{I.~I.}
  \bibnamefont{Tupitsyn}}, \bibinfo{journal}{Eur. Phys. J. D}
  \textbf{\bibinfo{volume}{38}}, \bibinfo{pages}{293} (\bibinfo{year}{2006}).

\bibitem[{\citenamefont{Volotka
  et~al.}(2008{\natexlab{a}})\citenamefont{Volotka, Glazov, Plunien, Shabaev,
  and Tupitsyn}}]{volotka:2008:167}
\bibinfo{author}{\bibfnamefont{A.~V.} \bibnamefont{Volotka}},
  \bibinfo{author}{\bibfnamefont{D.~A.} \bibnamefont{Glazov}},
  \bibinfo{author}{\bibfnamefont{G.}~\bibnamefont{Plunien}},
  \bibinfo{author}{\bibfnamefont{V.~M.} \bibnamefont{Shabaev}},
  \bibnamefont{and} \bibinfo{author}{\bibfnamefont{I.~I.}
  \bibnamefont{Tupitsyn}}, \bibinfo{journal}{Eur. Phys. J. D}
  \textbf{\bibinfo{volume}{48}}, \bibinfo{pages}{167}
  (\bibinfo{year}{2008}{\natexlab{a}}).

\bibitem[{\citenamefont{Indelicato et~al.}(2004)\citenamefont{Indelicato,
  Shabaev, and Volotka}}]{indelicato:2004:062506}
\bibinfo{author}{\bibfnamefont{P.}~\bibnamefont{Indelicato}},
  \bibinfo{author}{\bibfnamefont{V.~M.} \bibnamefont{Shabaev}},
  \bibnamefont{and} \bibinfo{author}{\bibfnamefont{A.~V.}
  \bibnamefont{Volotka}}, \bibinfo{journal}{Phys. Rev. A}
  \textbf{\bibinfo{volume}{69}}, \bibinfo{pages}{062506}
  (\bibinfo{year}{2004}).

\bibitem[{\citenamefont{{V.~M.~Shabaev, Izv. Vuz. Fiz. {\bf 33}, 43 (1990)
  [Sov. Phys. J. {\bf 33}, 660 (1990)].}}()}]{shabaev:1990:43}
\bibinfo{author}{\bibnamefont{{V.~M.~Shabaev, Izv. Vuz. Fiz. {\bf 33}, 43
  (1990) [Sov. Phys. J. {\bf 33}, 660 (1990)].}}}

\bibitem[{\citenamefont{{V.~M.~Shabaev, Teor. Mat. Fiz. {\bf 82}, 83 (1990)
  [Theor. Math. Phys. {\bf 82}, 57 (1990)].}}()}]{shabaev:1990:83}
\bibinfo{author}{\bibnamefont{{V.~M.~Shabaev, Teor. Mat. Fiz. {\bf 82}, 83
  (1990) [Theor. Math. Phys. {\bf 82}, 57 (1990)].}}}

\bibitem[{\citenamefont{Shabaev}(2002)}]{shabaev:2002:119}
\bibinfo{author}{\bibfnamefont{V.~M.} \bibnamefont{Shabaev}},
  \bibinfo{journal}{Phys. Rep.} \textbf{\bibinfo{volume}{356}},
  \bibinfo{pages}{119} (\bibinfo{year}{2002}).

\bibitem[{\citenamefont{{O.~Yu.~Andreev}
  et~al.}(2009)\citenamefont{{O.~Yu.~Andreev}, Labzowsky, and
  Plunien}}]{andreev:2009:032515}
\bibinfo{author}{\bibnamefont{{O.~Yu.~Andreev}}},
  \bibinfo{author}{\bibfnamefont{L.~N.} \bibnamefont{Labzowsky}},
  \bibnamefont{and} \bibinfo{author}{\bibfnamefont{G.}~\bibnamefont{Plunien}},
  \bibinfo{journal}{Phys. Rev. A} \textbf{\bibinfo{volume}{79}},
  \bibinfo{pages}{032515} (\bibinfo{year}{2009}).

\bibitem[{\citenamefont{{O.~Yu.~Andreev}
  et~al.}(2008)\citenamefont{{O.~Yu.~Andreev}, Labzowsky, Plunien, and
  Solovyev}}]{andreev:2008:135}
\bibinfo{author}{\bibnamefont{{O.~Yu.~Andreev}}},
  \bibinfo{author}{\bibfnamefont{L.~N.} \bibnamefont{Labzowsky}},
  \bibinfo{author}{\bibfnamefont{G.}~\bibnamefont{Plunien}}, \bibnamefont{and}
  \bibinfo{author}{\bibfnamefont{D.~A.} \bibnamefont{Solovyev}},
  \bibinfo{journal}{Phys. Rep.} \textbf{\bibinfo{volume}{455}},
  \bibinfo{pages}{135} (\bibinfo{year}{2008}).

\bibitem[{\citenamefont{Berestetsky et~al.}(1982)\citenamefont{Berestetsky,
  Lifshitz, and Pitaevsky}}]{berestetsky}
\bibinfo{author}{\bibfnamefont{V.~B.} \bibnamefont{Berestetsky}},
  \bibinfo{author}{\bibfnamefont{E.~M.} \bibnamefont{Lifshitz}},
  \bibnamefont{and} \bibinfo{author}{\bibfnamefont{L.~P.}
  \bibnamefont{Pitaevsky}}, \emph{\bibinfo{title}{Quantum Electrodynamics}}
  (\bibinfo{publisher}{{Pergamon Press, Oxford}}, \bibinfo{year}{1982}).

\bibitem[{\citenamefont{Itzykson and Zuber}(1980)}]{itzykson}
\bibinfo{author}{\bibfnamefont{C.}~\bibnamefont{Itzykson}} \bibnamefont{and}
  \bibinfo{author}{\bibfnamefont{J.-B.} \bibnamefont{Zuber}},
  \emph{\bibinfo{title}{Quantum Field Theory}}
  (\bibinfo{publisher}{{McGraw-Hill, New York}}, \bibinfo{year}{1980}).

\bibitem[{\citenamefont{Labzowsky et~al.}(2009)\citenamefont{Labzowsky,
  Solovyev, and Plunien}}]{labzowsky:2009:062514}
\bibinfo{author}{\bibfnamefont{L.}~\bibnamefont{Labzowsky}},
  \bibinfo{author}{\bibfnamefont{D.}~\bibnamefont{Solovyev}}, \bibnamefont{and}
  \bibinfo{author}{\bibfnamefont{G.}~\bibnamefont{Plunien}},
  \bibinfo{journal}{Phys. Rev. A} \textbf{\bibinfo{volume}{80}},
  \bibinfo{pages}{062514} (\bibinfo{year}{2009}).

\bibitem[{\citenamefont{Labzowsky and Shonin}(2004)}]{labzowsky:2004:012503}
\bibinfo{author}{\bibfnamefont{L.~N.} \bibnamefont{Labzowsky}}
  \bibnamefont{and} \bibinfo{author}{\bibfnamefont{A.~V.}
  \bibnamefont{Shonin}}, \bibinfo{journal}{Phys. Rev. A}
  \textbf{\bibinfo{volume}{69}}, \bibinfo{pages}{012503}
  (\bibinfo{year}{2004}).

\bibitem[{\citenamefont{Glazov et~al.}(2006)\citenamefont{Glazov, Volotka,
  Shabaev, Tupitsyn, and Plunien}}]{glazov:2006:330}
\bibinfo{author}{\bibfnamefont{D.~A.} \bibnamefont{Glazov}},
  \bibinfo{author}{\bibfnamefont{A.~V.} \bibnamefont{Volotka}},
  \bibinfo{author}{\bibfnamefont{V.~M.} \bibnamefont{Shabaev}},
  \bibinfo{author}{\bibfnamefont{I.~I.} \bibnamefont{Tupitsyn}},
  \bibnamefont{and} \bibinfo{author}{\bibfnamefont{G.}~\bibnamefont{Plunien}},
  \bibinfo{journal}{Phys. Lett. A} \textbf{\bibinfo{volume}{357}},
  \bibinfo{pages}{330} (\bibinfo{year}{2006}).

\bibitem[{\citenamefont{Volotka
  et~al.}(2008{\natexlab{b}})\citenamefont{Volotka, Glazov, Tupitsyn,
  Oreshkina, Plunien, and Shabaev}}]{volotka:2008:062507}
\bibinfo{author}{\bibfnamefont{A.~V.} \bibnamefont{Volotka}},
  \bibinfo{author}{\bibfnamefont{D.~A.} \bibnamefont{Glazov}},
  \bibinfo{author}{\bibfnamefont{I.~I.} \bibnamefont{Tupitsyn}},
  \bibinfo{author}{\bibfnamefont{N.~S.} \bibnamefont{Oreshkina}},
  \bibinfo{author}{\bibfnamefont{G.}~\bibnamefont{Plunien}}, \bibnamefont{and}
  \bibinfo{author}{\bibfnamefont{V.~M.} \bibnamefont{Shabaev}},
  \bibinfo{journal}{Phys. Rev. A} \textbf{\bibinfo{volume}{78}},
  \bibinfo{pages}{062507} (\bibinfo{year}{2008}{\natexlab{b}}).

\bibitem[{\citenamefont{Shabaev et~al.}(2004)\citenamefont{Shabaev, Tupitsyn,
  Yerokhin, Plunien, and Soff}}]{shabaev:2004:130405}
\bibinfo{author}{\bibfnamefont{V.~M.} \bibnamefont{Shabaev}},
  \bibinfo{author}{\bibfnamefont{I.~I.} \bibnamefont{Tupitsyn}},
  \bibinfo{author}{\bibfnamefont{V.~A.} \bibnamefont{Yerokhin}},
  \bibinfo{author}{\bibfnamefont{G.}~\bibnamefont{Plunien}}, \bibnamefont{and}
  \bibinfo{author}{\bibfnamefont{G.}~\bibnamefont{Soff}},
  \bibinfo{journal}{Phys. Rev. Lett.} \textbf{\bibinfo{volume}{93}},
  \bibinfo{pages}{130405} (\bibinfo{year}{2004}).

\bibitem[{\citenamefont{Angeli}(2004)}]{angeli:2004:185}
\bibinfo{author}{\bibfnamefont{I.}~\bibnamefont{Angeli}}, \bibinfo{journal}{At.
  Data Nucl. Data Tables} \textbf{\bibinfo{volume}{87}}, \bibinfo{pages}{185}
  (\bibinfo{year}{2004}).

\bibitem[{\citenamefont{Kozhedub et~al.}(2008)\citenamefont{Kozhedub, Andreev,
  Shabaev, Tupitsyn, Brandau, Kozhuharov, Plunien, and
  St\"{o}hlker}}]{kozhedub:2008:032501}
\bibinfo{author}{\bibfnamefont{Y.~S.} \bibnamefont{Kozhedub}},
  \bibinfo{author}{\bibfnamefont{O.~V.} \bibnamefont{Andreev}},
  \bibinfo{author}{\bibfnamefont{V.~M.} \bibnamefont{Shabaev}},
  \bibinfo{author}{\bibfnamefont{I.~I.} \bibnamefont{Tupitsyn}},
  \bibinfo{author}{\bibfnamefont{C.}~\bibnamefont{Brandau}},
  \bibinfo{author}{\bibfnamefont{C.}~\bibnamefont{Kozhuharov}},
  \bibinfo{author}{\bibfnamefont{G.}~\bibnamefont{Plunien}}, \bibnamefont{and}
  \bibinfo{author}{\bibfnamefont{T.}~\bibnamefont{St\"{o}hlker}},
  \bibinfo{journal}{Phys. Rev. A} \textbf{\bibinfo{volume}{77}},
  \bibinfo{pages}{032501} (\bibinfo{year}{2008}).

\bibitem[{\citenamefont{Artemyev et~al.}(2005)\citenamefont{Artemyev, Shabaev,
  Yerokhin, Plunien, and Soff}}]{artemyev:2005:062104}
\bibinfo{author}{\bibfnamefont{A.~N.} \bibnamefont{Artemyev}},
  \bibinfo{author}{\bibfnamefont{V.~M.} \bibnamefont{Shabaev}},
  \bibinfo{author}{\bibfnamefont{V.~A.} \bibnamefont{Yerokhin}},
  \bibinfo{author}{\bibfnamefont{G.}~\bibnamefont{Plunien}}, \bibnamefont{and}
  \bibinfo{author}{\bibfnamefont{G.}~\bibnamefont{Soff}},
  \bibinfo{journal}{Phys. Rev. A} \textbf{\bibinfo{volume}{71}},
  \bibinfo{pages}{062104} (\bibinfo{year}{2005}).

\bibitem[{\citenamefont{Dunford}(2004)}]{dunford:2004:062502}
\bibinfo{author}{\bibfnamefont{R.~W.} \bibnamefont{Dunford}},
  \bibinfo{journal}{Phys. Rev. A} \textbf{\bibinfo{volume}{69}},
  \bibinfo{pages}{062502} (\bibinfo{year}{2004}).

\bibitem[{\citenamefont{Simionovici et~al.}(1993)\citenamefont{Simionovici,
  Birkett, Briand, Charles, Dietrich, Finlayson, Indelicato, Liesen, and
  Marrus}}]{simionovici:1695:1993}
\bibinfo{author}{\bibfnamefont{A.}~\bibnamefont{Simionovici}},
  \bibinfo{author}{\bibfnamefont{B.~B.} \bibnamefont{Birkett}},
  \bibinfo{author}{\bibfnamefont{J.~P.} \bibnamefont{Briand}},
  \bibinfo{author}{\bibfnamefont{P.}~\bibnamefont{Charles}},
  \bibinfo{author}{\bibfnamefont{D.~D.} \bibnamefont{Dietrich}},
  \bibinfo{author}{\bibfnamefont{K.}~\bibnamefont{Finlayson}},
  \bibinfo{author}{\bibfnamefont{P.}~\bibnamefont{Indelicato}},
  \bibinfo{author}{\bibfnamefont{D.}~\bibnamefont{Liesen}}, \bibnamefont{and}
  \bibinfo{author}{\bibfnamefont{R.}~\bibnamefont{Marrus}},
  \bibinfo{journal}{Phys. Rev. A} \textbf{\bibinfo{volume}{48}},
  \bibinfo{pages}{1695} (\bibinfo{year}{1993}).

\bibitem[{\citenamefont{Sch\"afer et~al.}(1989)\citenamefont{Sch\"afer, Soff,
  Indelicato, M\"uller, and Greiner}}]{schaefer:1989:7362}
\bibinfo{author}{\bibfnamefont{A.}~\bibnamefont{Sch\"afer}},
  \bibinfo{author}{\bibfnamefont{G.}~\bibnamefont{Soff}},
  \bibinfo{author}{\bibfnamefont{P.}~\bibnamefont{Indelicato}},
  \bibinfo{author}{\bibfnamefont{B.}~\bibnamefont{M\"uller}}, \bibnamefont{and}
  \bibinfo{author}{\bibfnamefont{W.}~\bibnamefont{Greiner}},
  \bibinfo{journal}{Phys. Rev. A} \textbf{\bibinfo{volume}{40}},
  \bibinfo{pages}{7362} (\bibinfo{year}{1989}).

\bibitem[{\citenamefont{Labzowsky et~al.}(2001)\citenamefont{Labzowsky,
  Nefiodov, Plunien, Soff, Marrus, and Liesen}}]{labzowsky:2001:054105}
\bibinfo{author}{\bibfnamefont{L.~N.} \bibnamefont{Labzowsky}},
  \bibinfo{author}{\bibfnamefont{A.~V.} \bibnamefont{Nefiodov}},
  \bibinfo{author}{\bibfnamefont{G.}~\bibnamefont{Plunien}},
  \bibinfo{author}{\bibfnamefont{G.}~\bibnamefont{Soff}},
  \bibinfo{author}{\bibfnamefont{R.}~\bibnamefont{Marrus}}, \bibnamefont{and}
  \bibinfo{author}{\bibfnamefont{D.}~\bibnamefont{Liesen}},
  \bibinfo{journal}{Phys. Rev. A} \textbf{\bibinfo{volume}{63}},
  \bibinfo{pages}{054105} (\bibinfo{year}{2001}).
\end{thebibliography}
\end{document}